\documentclass[ALICE,manyauthors]{cernphprep}

\usepackage[comma,square,numbers,sort&compress]{natbib}
\usepackage[hidelinks]{hyperref}
\usepackage{lineno}
\usepackage[none]{hyphenat}
\usepackage{color}
\usepackage{capt-of}
\usepackage{array}
\usepackage[T1]{fontenc} 

\begin{document}%

\begin{titlepage}
\PHyear{2019} 
\PHnumber{287}      
\PHdate{Day Month}  
\PHdate{20 December}  
%

\title{Higher harmonic non-linear flow modes of charged hadrons in Pb--Pb collisions at $\sqrt{s_\mathrm{NN}}=$5.02 TeV}
\ShortTitle{Higher harmonic non-linear flow modes in Pb--Pb collisions at $\sqrt{s_\mathrm{NN}}=$5.02 TeV} 

\Collaboration{ALICE Collaboration\thanks{See Appendix~\ref{app:collab} for the list of collaboration members}}
\ShortAuthor{ALICE Collaboration} 

\begin{abstract}
Anisotropic flow coefficients, $v_n$, non-linear flow mode coefficients, $\chi_{n,mk}$, and correlations among different symmetry planes, $\rho_{n,mk}$ are measured in Pb--Pb collisions at $\sqrt{s_\mathrm{NN}}=5.02$ TeV. Results obtained with multi-particle correlations are reported for the transverse momentum interval $0.2<p_\mathrm{T}<5.0$ GeV/$c$ within the pseudorapidity interval $0.4<|\eta|<0.8$ as a function of collision centrality. The $v_n$ coefficients and $\chi_{n,mk}$ and $\rho_{n,mk}$ are presented up to the ninth and seventh harmonic order, respectively.
Calculations suggest that the correlations measured in different symmetry planes and the non-linear flow mode coefficients are dependent on the shear and bulk viscosity to entropy ratios of the medium created in heavy-ion collisions. The comparison between these measurements and those at lower energies and calculations from hydrodynamic models places strong constraints on the initial conditions and transport properties of the system. 
\end{abstract}
\end{titlepage}
\setcounter{page}{2}

%
%
\renewcommand{\floatpagefraction}{0.99}%
\section{Introduction}
One of the primary goals in the ultra-relativistic heavy-ion collision programs at the Relativistic Heavy Ion Collider (RHIC) and the Large Hadron Collider (LHC) is to study the nuclear matter at extreme conditions. The pressure gradients in the strongly interacting matter, known as the Quark--Gluon Plasma (QGP), are believed to drive the hydrodynamic expansion observed through anisotropy in multi-particle correlations in high energy collisions at RHIC and the LHC~\cite{Ollitrault:1992bk,Voloshin:2008dg}. The anisotropic expansion of the medium, commonly referred to as anisotropic flow~\cite{Ollitrault:1992bk}, can be characterized by a Fourier decomposition of the azimuthal particle distribution with respect to the reaction plane~\cite{Voloshin:1994mz,Poskanzer:1998yz} 
\begin{equation}
	\label{eq:fseriesexp}
	\frac{\mathrm{d}N}{\mathrm{d}\varphi}\propto 1+2\sum_{n=1}^{\infty}v_n\cos(n(\varphi-\psi_\mathrm{RP})),
\end{equation}
where the flow coefficient $v_n$ is the magnitude of the $n$-th order flow, and the reaction plane $\psi_\mathrm{RP}$ defined by the beam direction and impact parameter which is defined as the distance between the centers of two colliding nuclei.
Due to fluctuations in the initial state energy density profile, it is useful to define symmetry planes of different orders, where the $n$-th order plane $\psi_n$ defines
the orientation of the $n$-th order complex flow vector $V_n\equiv v_n\mathrm{e}^{in\psi_n}$. The expansion of the azimuthal distribution about $\psi_n$ also yields finite values of odd coefficients~\cite{Alver:2010gr,ALICE:2011ab}.
Anisotropic flow measurements through two- and multi-particle azimuthal correlations~\cite{Aamodt:2010pa,ALICE:2011ab,Abelev:2012di,Abelev:2014pua,Adam:2016nfo,Adam:2016izf,Acharya:2017ino,Acharya:2018zuq} have provided important information on the medium response and in particular its transport coefficients such as the shear viscosity to entropy density ratio ($\eta/s$), bulk viscosity to entropy density ratio ($\zeta/s$) 
and the equation of state~\cite{Teaney:2009qa}. Studies have shown the relativistic hydrodynamic nature of the medium~\cite{Ollitrault:1992bk,Molnar:2001ux,Teaney:2003kp,Lacey:2006bc,Drescher:2007cd,Xu:2007jv,Molnar:2008xj,Voloshin:2008dg,Heinz:2013th,Song:2017wtw}, with $\eta/s$ close to the AdS/CFT minimum $1/(4\pi)$~\cite{Kovtun:2004de}. 




The initial state eccentricity, determined from the energy density profile, is obtained from the definition~\cite{Alver:2010gr}
\begin{equation}
 \varepsilon_{n} \mathrm{e}^{in\Phi_{n}}=-\{r^n \mathrm{e}^{in\varphi}\}/\{r^n\},\,n\geq 2,
  \label{eq:eccentricities}
\end{equation}
where the curly brackets denote the average over the transverse plane, i.e. $\{\cdots\}$ = $\int \mathrm{d}x\mathrm{d}y\,e(x,y,\tau_0)(\cdots)$, $r$ is the distance to the system's center of mass, $\varphi$ is the azimuthal angle, $e(x,y,\tau_0)$ is the energy density at the initial time $\tau_0$, and $\Phi_{n}$ is the participant plane angle, defining the spatial symmetry of the nuclear constituents in the participant region (see Refs.~\cite{Teaney:2010vd,Niemi:2015qia}).
Hydrodynamic models demonstrate that the second and the third harmonic flow coefficients exhibit an almost linear dependence on the initial eccentricity coefficients $\varepsilon_n$~\cite{Niemi:2012aj}.
Considering that the anisotropic expansion is a result of a hydrodynamic evolution governed by $\eta/s$, a measurement of the second and third harmonics combined with hydrodynamic calculations can constrain the properties of the medium.
Several estimates for the limits of $\eta/s$ were determined through measurements of elliptic flow coefficient $v_2$~\cite{Adler:2001nb,Adler:2002pu,Adler:2003kt,Adams:2003zg,Adamczyk:2015obl,Alver:2007qw} and their comparison with hydrodynamic calculations. Consequently, the early constraints placed the value of $\eta/s$ between 0.08 to 0.16~\cite{Hirano:2005xf,Romatschke:2007mq,Chaudhuri:2007qp}.
However, the limited sensitivity of the elliptic flow to $\eta/s$ and the large uncertainty in the initial state anisotropy inhibit a precise determination of the value of $\eta/s$~\cite{Romatschke:2007mq,Song:2010mg,Luzum:2012wu,Shen:2011zc}, and its temperature dependence, which was recently shown to be explorable during the second run of LHC~\cite{ALICE:2016kpq,Acharya:2017gsw}.
In addition, part of the anisotropic flow can also originate from the hadronic phase~\cite{Bozek:2011ua,Rose:2014fba,Ryu:2015vwa}. It has been shown in \cite{Ryu:2015vwa,Ryu:2017qzn} that the inclusion of the temperature dependent bulk viscosity $\zeta/s$(T) in hydrodynamic simulation lead to a better description of the average transverse momentum of charged hadrons and on the elliptic flow coefficient. The effects of bulk viscosity should be considered when extracting any transport coefficient from the data~\cite{Bernhard:2016tnd,Dubla:2018czx,Bernhard2019}. 

Flow harmonics of order $n\geq 3$ reveal finer details of initial conditions~\cite{ALICE:2011ab,Abelev:2012di,Adam:2016nfo,Adam:2016izf,Acharya:2018zuq}, enabling to constrain $\eta/s$ better~\cite{Alver:2010dn,Aaboud:2018ves,ALICE:2016kpq,Acharya:2017gsw}.
Higher flow harmonics $n>3$ do not exhibit a linear response to the initial anisotropy \cite{Niemi:2012aj} as a finite contribution is induced by the initial state anisotropy of the lower orders~\cite{Gardim:2011xv,Gardim:2014tya}. For example, the fourth order flow vector $V_4$ gains contributions not only from the fourth order flow (linear flow mode), but also from the second order flow (non-linear flow mode).
Starting from the $V_n$ estimators studied in~\cite{Gardim:2011xv}, the flow can be expressed as a vector sum of the linear and non-linear modes
\begin{equation}
\label{eq:nldecomp}
\begin{split}
	V_4&=V_{4\mathrm{L}}+\chi_{4,22} V_2^2,\\
	V_5&=V_{5\mathrm{L}}+\chi_{5,23} V_2 V_3,\\
	V_6&=V_{6\mathrm{L}}+\chi_{6,222} V_2^3+\chi_{6,33} V_3^2+\chi_{6,24} V_2 V_{4L},\\
	V_7&=V_{7\mathrm{L}}+\chi_{7,223} V_2^2 V_3+\chi_{7,34} V_3 V_{4L}+\chi_{7,25} V_2 V_{5L},\\
	V_8&=V_{8\mathrm{L}}+\chi_{8,2222} V_2^4+\chi_{8,233} V_2 V_3^2+\mathcal{E}(V_{4L},V_{5L},V_{6L}),
\end{split}
\end{equation}
where $\chi_{n,mk}$ is called non-linear flow mode coefficient, characterizing the non-linear flow mode induced by the lower order harmonics. The high order linear component is denoted by $V_{n\mathrm{L}}$, while the many higher order linear couplings are depicted by $\mathcal{E}(\dots)$ for $V_8$. The $V_{n\mathrm{L}}$ is linearly related to a cumulant-defined anisotropy~\cite{Teaney:2013dta}
\begin{equation}
\varepsilon_4'\mathrm{e}^{i4\Phi_4'}=\varepsilon_4\mathrm{e}^{i4\Phi_4}+\frac{3\langle r^2\rangle^2}{\langle r^4\rangle}\varepsilon_2\mathrm{e}^{i4\Phi_2}
\end{equation}
as opposed to the relation $v_n\propto\varepsilon_n$, where $v_n$ is the magnitude of the total contribution and $\varepsilon_n$ is given by Eq.~(\ref{eq:eccentricities}).
In earlier measurements performed by ALICE~\cite{Acharya:2017zfg}, the non-linear flow mode coefficients were measured up to the sixth harmonic order in Pb--Pb collisions at $\sqrt{s_\mathrm{NN}}=2.76\,\mathrm{TeV}$. It was indicated that the coefficients $\chi_{5,23}$ and $\chi_{6,33}$ are sensitive not only to $\eta/s$, but also to the distinctive energy density profiles generated by different initial conditions. It was observed that the hydrodynamic models with their respective initial conditions Monte-Carlo (MC)-Glauber~\cite{Miller:2007ri,Qiu:2011iv}, MC-KLN~\cite{Hirano:2005xf,Drescher:2007ax}, and IP-Glasma~\cite{McDonald:2016vlt}), are unable to reproduce these measurements, which indicates that the model tuning and $\eta/s$ parameterization require further work.

In this paper, measurements of high order flow coefficients in Pb--Pb collisions at $\sqrt{s_\mathrm{NN}}=5.02\,\mathrm{TeV}$ are presented. The flow coefficients $v_n$ are measured up to the ninth harmonic, $v_9$, extending the previous measurements of $v_2$--$v_6$~\cite{Acharya:2018lmh}. The data recorded during the 2015 heavy-ion run of the LHC allow the measurements of non-linear flow mode and correlations between symmetry planes to be extended. A total of six non-linear flow mode coefficients are measured, including the non-linear flow mode coefficient $\chi_{7,223}$, for which the sensitivity to $\eta/s$ is expected to be significantly stronger than for the lower odd-harmonic coefficient $\chi_{5,23}$~\cite{Luzum:2012wu,Yan:2015jma}. The results are compared with those in Pb--Pb collisions at $\sqrt{s_\mathrm{NN}}=2.76\,\mathrm{TeV}$~\cite{Acharya:2017zfg} and various state of the art hydrodynamical calculations. 



\section{Formalism and Observables}
\label{sec:obs}
In order to separate the linear and non-linear contributions from Eq.~(\ref{eq:nldecomp}), one assumes the respective contributions to be uncorrelated~\cite{Teaney:2012ke}. For example for the fourth order $V_4$, by mean-squaring the equations in Eq.~(\ref{eq:nldecomp}) and setting $\langle (V_2^*)^2 V_{4L}\rangle\simeq\langle V_2^2 V_{4L}^*\rangle\simeq 0$, the linear part can be derived
\begin{equation}
	\label{eq:linear}
	\underbrace{\langle|V_{4L}|^2\rangle^{\frac{1}{2}}}_{v_{4\mathrm{L}}}=(\underbrace{\langle|V_4|^2\rangle}_{v_4^2}-\underbrace{\chi_{4,22}^2\langle |V_2|^4\rangle}_{v_{4,\mathrm{NL}}^2})^{\frac{1}{2}}.
\end{equation}
Here $\langle\rangle$ denotes an average over all events and ${}^*$ the complex conjugate. The magnitudes of the linear and non-linear flow coefficients are denoted with $v_{4\mathrm{L}}$ and $v_{4,\mathrm{NL}}$, respectively.

The observables of the non-linear response mode are constructed from the projections of flow vectors on to the symmetry planes of lower harmonics~\cite{Jia:2012ma,Luzum:2011mm}. 
For $n=4$, the magnitude of the non-linear response mode is given by
\begin{equation}
\label{eq:proj}
	v_{4,22}=\frac{\Re\langle V_4 (V_2^*)^2\rangle}{\sqrt{\langle|V_2|^4\rangle}}\approx\langle v_4\cos(4\psi_4-4\psi_2)\rangle,
\end{equation}
where $v_{4,22}^2\equiv v_{4,\mathrm{NL}}^2\equiv\chi_{4,22}^2\langle|V_2|^4\rangle$. The right-hand side approximation holds if the low ($n=2,3$) and high order flow is weakly correlated.
Only the fourth harmonic is shown here and the complete list of other harmonics are provided in Appendix A.

The contributions from short-range correlations unrelated to the common symmetry plane, commonly referred to as ``non-flow'', are suppressed by using the subevent method where the event is divided into two subevents separated by a pseudorapidity gap~\cite{Poskanzer:1998yz}. The underlying multi-particle correlation coefficient for subevent A is $v_{4,22}^\mathrm{A}=\langle\langle\cos(4\varphi_1^\mathrm{A}-2\varphi_2^\mathrm{B}-2\varphi_3^\mathrm{B})\rangle\rangle\allowbreak{}/\langle\langle\cos(2\varphi_1^\mathrm{A}+2\varphi_2^\mathrm{A}-2\varphi_3^\mathrm{B}-2\varphi_4^\mathrm{B}\rangle\rangle^{1/2}$ as determined using Eq.~(\ref{eq:proj}),\footnote{For practical usage, the self-correlation is recursively removed from three- and four-particle correlations, resulting in modified equations.} and 
a similar treatment is applied for the subevent B, for which $v_{4,22}^\mathrm{B}$ is obtained by swapping B for A in the aforegiven expression. The final result is then the average of the results from subevents A and B.

The symmetry-plane correlations are defined as the ratio between the magnitude of the non-linear flow modes and flow coefficients \cite{Qiu:2012uy}. For $n=4$, one obtains
\begin{equation}
	\rho_{4,22}=\frac{v_{4,22}}{v_4}\approx\langle\cos(4\psi_4-4\psi_2)\rangle.
\end{equation}
A value close to zero indicates weakly correlated symmetry planes, while a value reaching one implies a strong correlation. The correlations between symmetry planes reflect those of the corresponding initial state participant planes~~\cite{Aad:2014fla,Acharya:2017zfg}, therefore providing valuable information on the evolution of the QGP. 
Correlations between symmetry planes have been previously studied using the event-plane method~\cite{Aad:2014fla,Chatrchyan:2013kba}, \emph{event plane} describing an experimentally approximated symmetry plane. However, these results depend on the event-plane
res\-{}o\-{}lu\-{}tion~\cite{Luzum:2012da}, which complicates the comparison between data and theoretical calculations.
Recently, the ALICE Collaboration has measured symmetry-plane correlations~\cite{Acharya:2017zfg}.
It was found that the correlations of symmetry planes of higher harmonics with second and third order symmetry planes increased for less central collisions. Furthermore, the comparison with hydrodynamic calculations revealed the importance of final-state collective dynamics in addition to the initial-state density fluctuations~\cite{Hirano:2005xf} as it is known that the observation of correlated final state symmetry planes implies the existence of fluctuations in the initial state eccentricity vectors.

The fourth non-linear flow mode coefficient, with the aforementioned assumptions, is given by~\cite{Yan:2015jma}
\begin{equation}
\label{eq:chi}
	\chi_{4,22}=\frac{v_{4,22}}{\sqrt{\langle v_2^4\rangle}}.
\end{equation}




\section{Experimental Setup and Data Analysis}
\label{sec:analysis}
The data sample consists of about 42 million minimum bias Pb--Pb collisions at $\sqrt{s_\mathrm{NN}}=5.02\,\mathrm{TeV}$, recorded by ALICE~\cite{Aamodt:2008zz,Abelev:2014ffa} during the 2015 heavy-ion run at the LHC. Detailed descriptions of the detector can be found in~\cite{Aamodt:2008zz,Carminati:2004fp,Alessandro:2006yt}. The trigger plus crossing of beam is provided by signals from the two scintillator arrays, V0A and V0C~\cite{Aamodt:2008zz,Abbas:2013taa}, covering the pseudorapidity intervals $2.8<\eta<5.1$ and $-3.7<\eta<-1.7$, respectively. A primary vertex position less than 10 cm in beam direction from the nominal interaction point is required. Pile-up events are removed by correlating the V0 multiplicity with the multiplicity from the first Silicon Pixel Detector (SPD)~\cite{Aamodt:2008zz,Aamodt:2010aa} layer. To further remove pile-up events, information from the Time-of-Flight (TOF)~\cite{Carnesecchi:2018oss} detector is used: 
the multiplicity estimates from the SPD are correlated with those imposed with a TOF readout requirement. The centrality of the collision is determined using information from the V0 arrays. Further details on the centrality determination in ALICE are given in~\cite{Abelev:2013qoq}. Only events in the centrality range 0\% to 60\% are used in the analysis. 

The track reconstruction is based on combined information from the Time Projection Chamber (TPC)~\cite{Aamodt:2008zz,Alme:2010ke} and the Inner Tracking System (ITS)~\cite{Aamodt:2008zz,Aamodt:2010aa}. To avoid contributions from secondary particles, the tracks are required to have a distance of closest approach to the primary vertex of less than 3.2 cm and 2.4 cm in the longitudinal and transverse directions, respectively. Such a loose Distance of Closest Approach (DCA) track cut is chosen to improve the uniformity of the $\varphi$-distribution for the $Q_n$-vector calculation. Furthermore, each track is required to have at least 70 TPC space points out of the maximum 159, and the average $\chi^2$ per degree of freedom of the track fit to the TPC space points to be less than 2. Minimum 2 hits are required in the ITS. In order to counteract the effects of track reconstruction efficiency and contamination from secondary particles~\cite{ALICE-PUBLIC-2017-005}, a HIJING simulation~\cite{Wang:1991hta,Gyulassy:1994ew} with GEANT3~\cite{Brun:1994aa} detector model is employed to construct a $p_\mathrm{T}$-dependent track weighting correction. The track reconstruction efficiency is approximately 65\% at $p_T=0.2\,\mathrm{GeV}/c$ and 80\% at $p_T>1.0\,\mathrm{GeV}/c$, while the contamination from secondaries is less than 10\% and 5\%, respectively. Only particle tracks within the transverse momentum interval $0.2<p_T<5.0\,\mathrm{GeV}/c$ and pseudorapidity range $0.4<|\eta|<0.8$ are considered. A pseudorapidity gap $|\Delta\eta|>0.8$ is used to suppress the non-flow. The observables in this analysis are measured with multi-particle correlations obtained using the generic framework for anisotropic flow analysis~\cite{Bilandzic:2013kga}.


\section{Systematic Uncertainties}

The systematic uncertainties are estimated by varying criteria for selecting the events and tracks. The systematic evaluation is done by independently varying the selection criteria, and the results given by this variation are then compared to the default criteria given in Sec.~\ref{sec:analysis}. The total uncertainty is obtained by assuming that the individual sources are uncorrelated, which are then quadratically summed.

Summaries of the relative systematic uncertainties are given in Tabs.~\ref{tab:vnsyst}--\ref{tab:chisyst}. Uncertainties stemming from the event selection criteria are estimated by changing the rejection based on the vertex position from 10 cm to 8 cm and by adjusting the pile-up rejection criteria. It is found that the contribution to the uncertainty is generally negligible, below 1\%. An alternative centrality determination is employed using the event multiplicity estimates from the SPD layers. 
The uncertainty related to the centrality determination is less than 2\% for all observables, except for $v_7$ to $v_9$ for which the uncertainty increases to 10\%.

\makeatletter
\newcommand{\thickhline}{%
    \noalign {\ifnum 0=`}\fi \hrule height 1pt
    \futurelet \reserved@a \@xhline
}
\newcolumntype{"}{@{\hskip\tabcolsep\vrule width 1pt\hskip\tabcolsep}}
\makeatother

\begin{table}[h!]
	\caption{Relative systematic uncertainties of the flow coefficients. The uncertainties are given in percents and are categorized into four groups: event selection, centrality determination, tracking and non-flow. The overall systematic uncertainty is obtained by summing in quadrature the uncertainties from each source.} 
	\centering
	{\small
	\begin{tabular}{|c|cccccccc|}
		\thickhline
		Type & $v_2$ & $v_3$ & $v_4$ & $v_5$ & $v_6$ & $v_7$ & $v_8$ & $v_9$\\
\thickhline\multicolumn{9}{|c|}{Event Selection}\\ \hline
$z$-vertex cut & $<$ 0.1 & $<$ 0.1 & $<$ 0.1 & 0.5 & 1.2 & 1.6 & 1.8 & 1.7\\
Pile-up rejection & $<$ 0.1 & $<$ 0.1 & $<$ 0.1 & 0.2 & 0.8 & 1.3 & 1.7 & 2.0\\
\hline\multicolumn{9}{|c|}{Centrality Determination}\\ \hline
SPD & 0.6 & 0.3 & 0.3 & 1.1 & 3.9 & 6.6 & 9.1 & 11.5\\
\hline\multicolumn{9}{|c|}{Tracking}\\ \hline
Magnetic field polarity & 0.1 & 0.1 & 1.7 & 2.4 & 4.1 & 6.8 & 10.5 & 15.2\\
Tracking mode & 0.1 & 0.2 & $<$ 0.1 & 2.4 & 5.4 & 7.2 & 7.6 & 6.8\\
Number of TPC space points & 0.7 & 1.2 & 1.4 & 1.5 & 1.6 & 1.7 & 1.7 & 1.8\\
Space charge distortion & $<$ 0.1 & $<$ 0.1 & $<$ 0.1 & 0.2 & 0.7 & 1.2 & 1.7 & 2.3\\
\hline\multicolumn{9}{|c|}{Non-flow}\\ \hline
Charge combinations ($--$/$++$) & 1.1 & 0.7 & 0.8 & 2.9 & 6.2 & 9.3 & 12.3 & 15.2\\
\hline\multicolumn{9}{|c|}{}\\ \hline
Overall & 1.5 & 1.4 & 2.4 & 4.9 & 10.3 & 15.4 & 20.4 & 25.6\\
		\hline
	\end{tabular}
	}
	\label{tab:vnsyst}
\end{table}

\begin{table}[h!]
	\caption{Relative systematic uncertainties of the harmonic projections $v_{n,mk}$.}
	\centering
	{\small
	\begin{tabular}{|c|cccccc|}
		\thickhline
		Type & $v_{4,22}$ & $v_{5,23}$ & $v_{6,222}$ & $v_{6,33}$ & $v_{6,24}$ & $v_{7,223}$\\ 
\thickhline\multicolumn{7}{|c|}{Event Selection}\\ \hline
$z$-vertex cut & 0.1 & 0.1 & 0.2 & 0.3 & 0.2 & 0.1\\
Pile-up rejection & $<$ 0.1 & 0.1 & 0.4 & 0.5 & 0.4 & $<$ 0.1\\
\hline\multicolumn{7}{|c|}{Centrality Determination}\\ \hline
SPD & 1.5 & 0.7 & 0.3 & 0.3 & 0.7 & 1.4\\
\hline\multicolumn{7}{|c|}{Tracking}\\ \hline
Magnetic field polarity & 0.5 & 0.5 & 1.9 & 3.2 & 4.4 & 5.5\\
Tracking mode & 0.1 & 0.4 & 1.4 & 1.7 & 1.1 & $<$ 0.1\\
Number of TPC space points & 3.8 & 2.3 & 1.5 & 1.4 & 2.1 & 3.5\\
Space charge distortion & 0.2 & 0.1 & 1.8 & 4.0 & 6.7 & 9.9\\
\hline\multicolumn{7}{|c|}{Non-flow}\\ \hline
Charge combinations ($--$/$++$) & 4.2 & 4.7 & 5.8 & 7.4 & 9.6 & 14.3\\
\hline\multicolumn{7}{|c|}{}\\ \hline
Overall & 5.9 & 5.3 & 6.7 & 9.3 & 12.7 & 18.6\\
		\hline
	\end{tabular}
	}
	\label{tab:projsyst}
\end{table}

\begin{table}[h!]
	\caption{Relative systematic uncertainties of the symmetry-plane correlations $\rho_{n,mk}$.}
	\centering
	{\small
	\begin{tabular}{|c|cccccc|}
		\thickhline
		Type & $\rho_{4,22}$ & $\rho_{5,23}$ & $\rho_{6,222}$ & $\rho_{6,33}$ & $\rho_{6,24}$ & $\rho_{7,223}$\\
\thickhline\multicolumn{7}{|c|}{Event Selection}\\ \hline
$z$-vertex cut & 0.1 & 0.3 & 0.1 & 0.2 & 0.8 & 2.5\\
Pile-up rejection & 0.1 & 0.3 & 0.1 & 0.3 & 1.0 & 2.2\\
\hline\multicolumn{7}{|c|}{Centrality Determination}\\ \hline
SPD & 0.9 & 0.3 & 0.7 & 0.9 & 1.2 & 1.5\\
\hline\multicolumn{7}{|c|}{Tracking}\\ \hline
Magnetic field polarity & $<$ 0.1 & 1.8 & 6.8 & 10.1 & 13.8 & 18.0\\
Tracking mode & 0.1 & 0.3 & 0.8 & 2.6 & 6.1 & 11.2\\
Number of TPC space points & $<$ 0.1 & 0.7 & 0.1 & $<$ 0.1 & 1.0 & 2.8\\
Space charge distortion & 0.2 & 0.2 & 1.5 & 3.5 & 6.7 & 11.1\\
\hline\multicolumn{7}{|c|}{Non-flow}\\ \hline
Charge combinations ($--$/$++$) & 3.1 & 3.6 & 3.6 & 5.6 & 8.7 & 12.9\\
\hline\multicolumn{7}{|c|}{}\\ \hline
Overall & 3.3 & 4.2 & 7.9 & 12.4 & 18.8 & 27.5\\
		\hline
	\end{tabular}
	}
	\label{tab:rhosyst}
\end{table}

\begin{table}[h!]
	\caption{Relative systematic uncertainties of the non-linear flow mode coefficients $\chi_{n,mk}$.}
	\centering
	{\small
	\begin{tabular}{|c|cccccc|}
		\thickhline
		Type & $\chi_{4,22}$ & $\chi_{5,23}$ & $\chi_{6,222}$ & $\chi_{6,33}$ & $\chi_{6,224}$ & $\chi_{7,223}$ \\
\thickhline\multicolumn{7}{|c|}{Event Selection}\\ \hline
$z$-vertex cut & $<$ 0.1 & 0.1 & 0.3 & 0.3 & 0.3 & 0.1\\
Pile-up rejection & $<$ 0.1 & 0.1 & 0.5 & 0.6 & 0.5 & 0.1\\
\hline\multicolumn{7}{|c|}{Centrality Determination}\\ \hline
SPD & 0.2 & 0.6 & 1.0 & 1.0 & 0.7 & 0.1\\
\hline\multicolumn{7}{|c|}{Tracking}\\ \hline
Magnetic field polarity & 0.6 & 0.2 & 2.5 & 4.1 & 5.1 & 5.5\\
Tracking mode & $<$ 0.1 & 0.2 & 1.4 & 1.7 & 1.2 & 0.2\\
Number of TPC space points & $<$ 0.1 & 0.2 & 0.5 & 0.7 & 0.9 & 1.1\\
Space charge distortion & 0.2 & 0.1 & 1.9 & 4.4 & 7.1 & 10.1\\
\hline\multicolumn{7}{|c|}{Non-flow}\\ \hline
Charge combinations ($--$/$++$) & 0.2 & 1.5 & 7.7 & 12.0 & 14.4 & 15.0\\
\hline\multicolumn{7}{|c|}{}\\ \hline
Overall & 0.7 & 1.7 & 8.5 & 13.6 & 17.0 & 19.0\\
		\hline
	\end{tabular}
	}
	\label{tab:chisyst}
\end{table}

The ALICE detector can be operated with either positive or negative solenoid magnetic field polarity. The polarity of the field affects the direction of the charged particle curvature, while also subjecting the structural materials of the detector itself to either a positive or negative magnetic field. The default data set is composed of events recorded with both polarities. The results produced with exclusively either negative or positive magnetic field configurations deviate from the default by up to 15\% in case of flow coefficients, and 28\% for $\rho_{7,223}$.
In order to estimate the non-flow contributions from resonance decays, the like-sign technique~\cite{Voloshin:2008dg} which correlates exclusively either positively or negatively charged particles, is employed. The difference with respect to the results obtained by selecting all charged particles is assigned as a systematic uncertainty. In general, this uncertainty ranges from 2\% to 15\%.
The effect from the space charge distortions in the TPC drift volume because of the higher interaction rates is estimated by comparing results from different regions of the TPC, one for $\eta>0$ and the other $\eta<0$. The maximum uncertainty is evaluated less than 15\%. The track reconstruction related uncertainty, referred to as tracking mode, is evaluated by comparing the results obtained with tracks for which the requirement for the number of hits in the ITS layers is changed. 
In this case, the uncertainty is generally less than 15\%, and a maximum 20\% is evaluated for $\rho_{7,223}$. Furthermore, the track selection criteria is tightened by increasing the minimum number of the TPC space points from 70 to 90, resulting in uncertainties around 1\% to 3\%.

\section{Results}
\newcommand{\trento}{T\raisebox{-.5ex}{R}ENTo}

\begin{figure}[!tbp]
	\centering
	\includegraphics[width=1.0\textwidth]{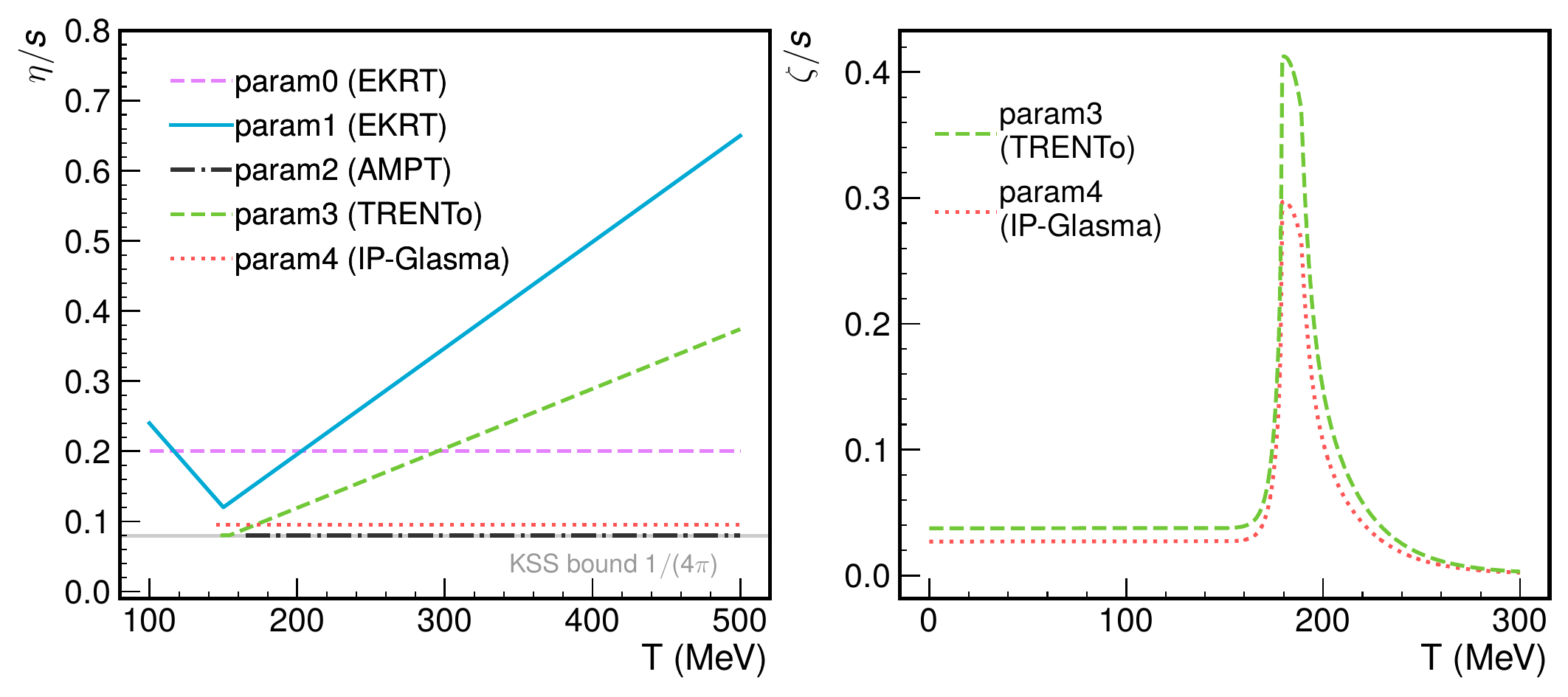}
	\caption{The five different parameterizations of $\eta/s$ and $\zeta/s$ used for the different hydrodynamic model calculations are shown in the left and right panel. Note that the functional form of $\zeta/s(T)$ is the same for param3 and param4 and taken from Eq.~5 in \cite{Bernhard:2016tnd} motivated by Refs.~\cite{Ryu:2015vwa,Denicol:2009am,Karsch:2007jc,NoronhaHostler:2008ju}. For the parameters with \trento{} initial condition, the ones based on identified yields are taken from Table~4 in \cite{Bernhard:2016tnd}. The $\zeta/s$ normalization factor used with IP-Glasma (\trento) initial conditions is 0.9 (1.25). The models with $\zeta/s=0$ are not shown on the right.}
 	\label{fig:params}

	\captionof{table}{Hydrodynamic model configurations. Shown are the key components such as initial condition models, and $\eta/s$ and $\zeta/s$ parameterizations. With \trento{} initial conditions, an entropy deposition parameter $p=0$~\cite{Zhao:2017yhj} is used for all calculations.}
	\centering
	{\small
	\begin{tabular}{|c|c|c|c|c|c|}
		\hline
		Model & Hydrodynamic code & Initial conditions & $\eta/s$ & $\zeta/s$\\
		\hline
		{EKRT+param0}~\cite{Niemi:2015qia,Niemi:2015voa} & {EbyE}~\cite{Niemi:2015qia,Molnar:2009tx} & EKRT~\cite{Niemi:2015qia,Niemi:2015voa} & 0.20 & 0 \\
		{EKRT+param1}~\cite{Niemi:2015qia,Niemi:2015voa} & {EbyE}~\cite{Niemi:2015qia,Molnar:2009tx} & EKRT~\cite{Niemi:2015qia,Niemi:2015voa} & $\eta/s(T)$~\cite{Niemi:2015qia} & 0 \\
		{AMPT+param2}~\cite{Zhao:2017yhj} & {iEBE-VISHNU}~\cite{Shen:2014vra} & AMPT~\cite{Bhalerao:2015iya,Pang:2012he,Xu:2016hmp} & 0.08 & 0 \\
		{\trento+param3}~\cite{Zhao:2017yhj} & {iEBE-VISHNU}~\cite{Shen:2014vra} & \trento($p=0$)~\cite{Moreland:2014oya} & $\eta/s(T)$~\cite{Zhao:2017yhj,Bernhard:2016tnd} & $\zeta/s(T)$~\cite{Zhao:2017yhj,Bernhard:2016tnd} \\
		{IP-Glasma+param4}~\cite{McDonald:2016vlt} & MUSIC~\cite{Schenke:2010rr} & IP-Glasma~\cite{Schenke:2012wb} & 0.095 & $\zeta/s(T)$~{\cite{McDonald:2016vlt}} \\
		\hline
	\end{tabular}}
	\label{tab:hydro}
\end{figure}

In this section, the measurements of the flow coefficients, the non-linear modes, symmetry-plane corre-\allowbreak{}lations and the non-linear flow mode coefficients are presented. They are compared with hydrodynamic calculations with various settings~\cite{Niemi:2015qia,Niemi:2015voa,Zhao:2017yhj,McDonald:2016vlt}. 
The first calculation is based on an event-by-event viscous hydrodynamic model with EKRT initial conditions~\cite{Niemi:2015qia,Niemi:2015voa} using a value of $\eta/s=0.20$ ({param0}) and a temperature dependent $\eta/s(T)$ ({param1}). For both configurations, $\zeta/s$ is set to zero. The visualization of the model parameters can be found in Fig.~\ref{fig:params}. The second calculation employs the {iEBE-VISHNU} hybrid model~\cite{Shen:2014vra} with AMPT~\cite{Bhalerao:2015iya,Pang:2012he,Xu:2016hmp} and \trento~\cite{Moreland:2014oya} initial conditions. The $\eta/s=0.08$ and $\zeta/s=0$ are taken for {param2}, while the $\eta/s(T)$ and $\zeta/s(T)$ ({param3}), extracted using Bayesian analysis~\cite{Bernhard:2016tnd} (except for the normalization factors) from a fit to the final multiplicities of the charged hadron spectra in Pb--Pb collisions at $\sqrt{s_\mathrm{NN}}=5.02$, are used for the \trento{} initial conditions. 
The third calculation uses the MUSIC model~\cite{Schenke:2010rr} with IP-Glasma~\cite{Schenke:2012wb} initial conditions with a value of $\eta/s=0.095$ and $\zeta/s(T)$ ({param4}). Each of the $\eta/s(T)$ parameterizations is adjusted to reproduce the measured charged hadron multiplicity, the low-$p_\mathrm{T}$ region of the charged-hadron spectra, and $v_n$ from central to mid-peripheral collisions up to the fourth harmonic at RHIC and the LHC~\cite{Niemi:2015qia,Qiu:2011hf,Shen:2010uy,Shen:2011eg,Bhalerao:2015iya,Ryu:2017qzn,McDonald:2016vlt}. 
The model configurations are summarized in Tab.~\ref{tab:hydro}.

\begin{figure}[!tbp]
	\centering
	\includegraphics[width=1.0\textwidth]{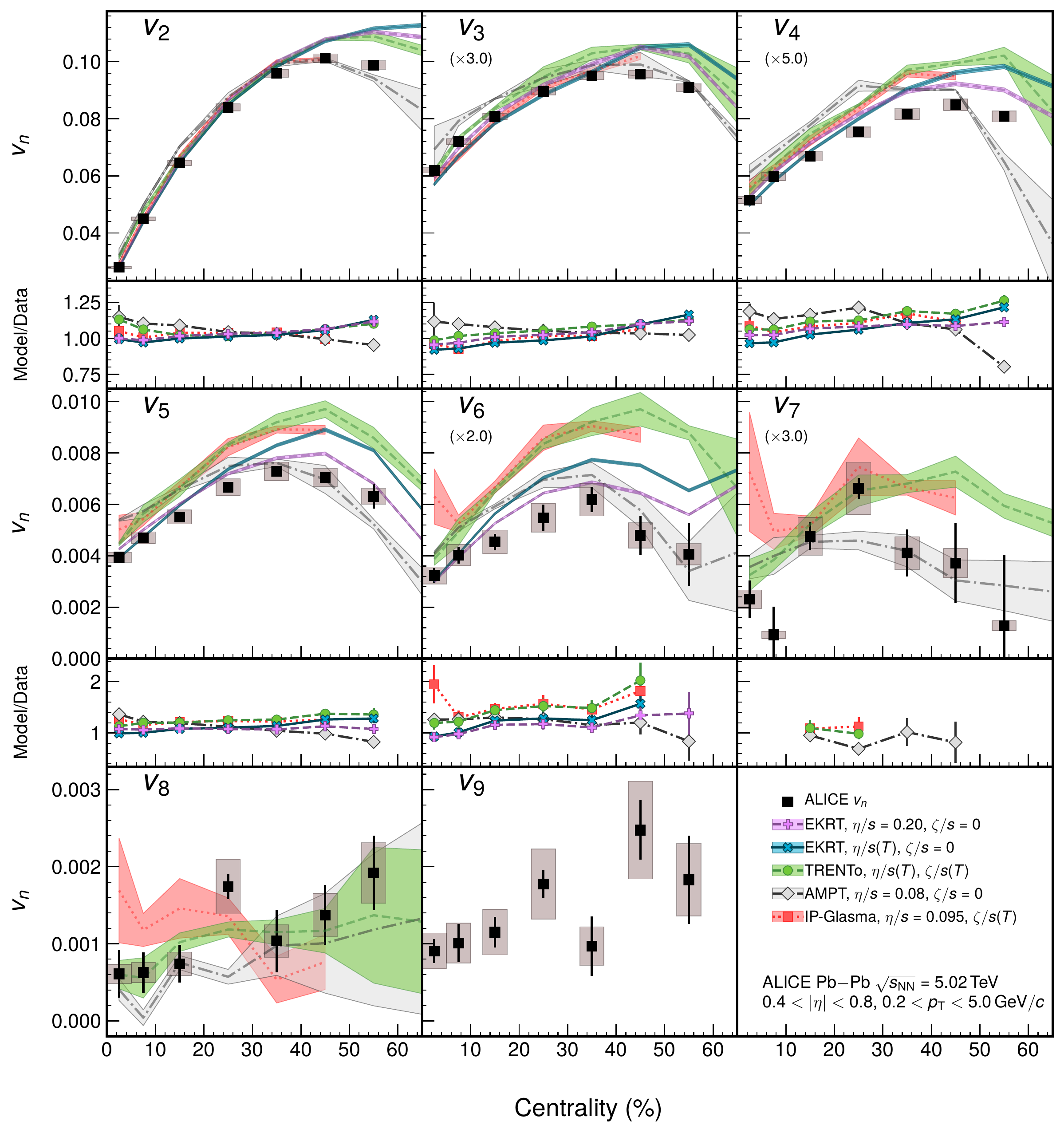}
	\caption{Flow harmonics up to the ninth order as a function of centrality, along with five different hydrodynamic calculations shown as color bands, each representing different configurations. For the black markers representing the measured data points, the sytematic uncertainty is indicated by the gray patches around the markers. The bands indicate the extent of the uncertainty of the corresponding calculation. On the bottom part of each panel, the ratios between model calculations and the data are shown with symbols. Ratios with uncertainties larger than 1 are not shown in the ratio panel. For some panels, the points are scaled by an indicated factor for better visibility across the panels.} 
 	\label{fig:vns}
\end{figure}

In Fig.~\ref{fig:vns}, the measurements of the flow coefficients from $v_2$ to $v_9$ are presented. The first two coefficients up to $v_6$ have been extensively measured at RHIC and LHC~\cite{Aamodt:2010pa,ALICE:2011ab,Abelev:2012di,Abelev:2014pua,Adam:2016nfo,Adam:2016izf,Acharya:2017ino,Acharya:2018zuq}, and more recently also $v_7$~\cite{Aaboud:2018ves}. 
The present analysis reports the first results on higher harmonic coefficients from $v_7$ to $v_9$ in ALICE, where $v_8$ and $v_9$ are measured for the first time at the LHC energies.
The coefficients exhibit a modest centrality dependence, and their magnitude is similar to that of $v_7$ within statistical and systematic uncertainties. The measurements up to $v_6$ are compatible with those published previously~\cite{Acharya:2018lmh}.

Figure~\ref{fig:vns} also shows the comparison between the measured $v_n$ and model calculations. The hydrodynamic calculations qualitatively reproduce the $v_n$ measurements, and the overall model depiction is very good for $v_2$ and $v_3$. For $n\geq 4$ however, the calculations show noticeable overestimations, especially in mid-peripheral collisions. For $v_5$ and $v_6$, the data are well described by {EKRT+\allowbreak{}param0}, showing a better agreement than the temperature dependent {EKRT+\allowbreak{}param1}. The data are also described by {AMPT+\allowbreak{}param2}, for which the agreement for $v_5$ and $v_6$ is good in mid-central and mid-peripheral collisions. {IP-Glasma+\allowbreak{}param4} and {\trento+\allowbreak{}param3} overestimate the measurements by a factor of 1.5$\sim$2. Values of $v_7$ are well estimated by {AMPT+\allowbreak{}param2}, and $v_8$ by both {AMPT+\allowbreak{}param2} and {\trento+\allowbreak{}param3} within uncertainties. In other cases, the data are overestimated by the other models. 

\begin{figure}[tbp]
	\centering
	\hspace{-3em}\includegraphics[width=0.90\textwidth]{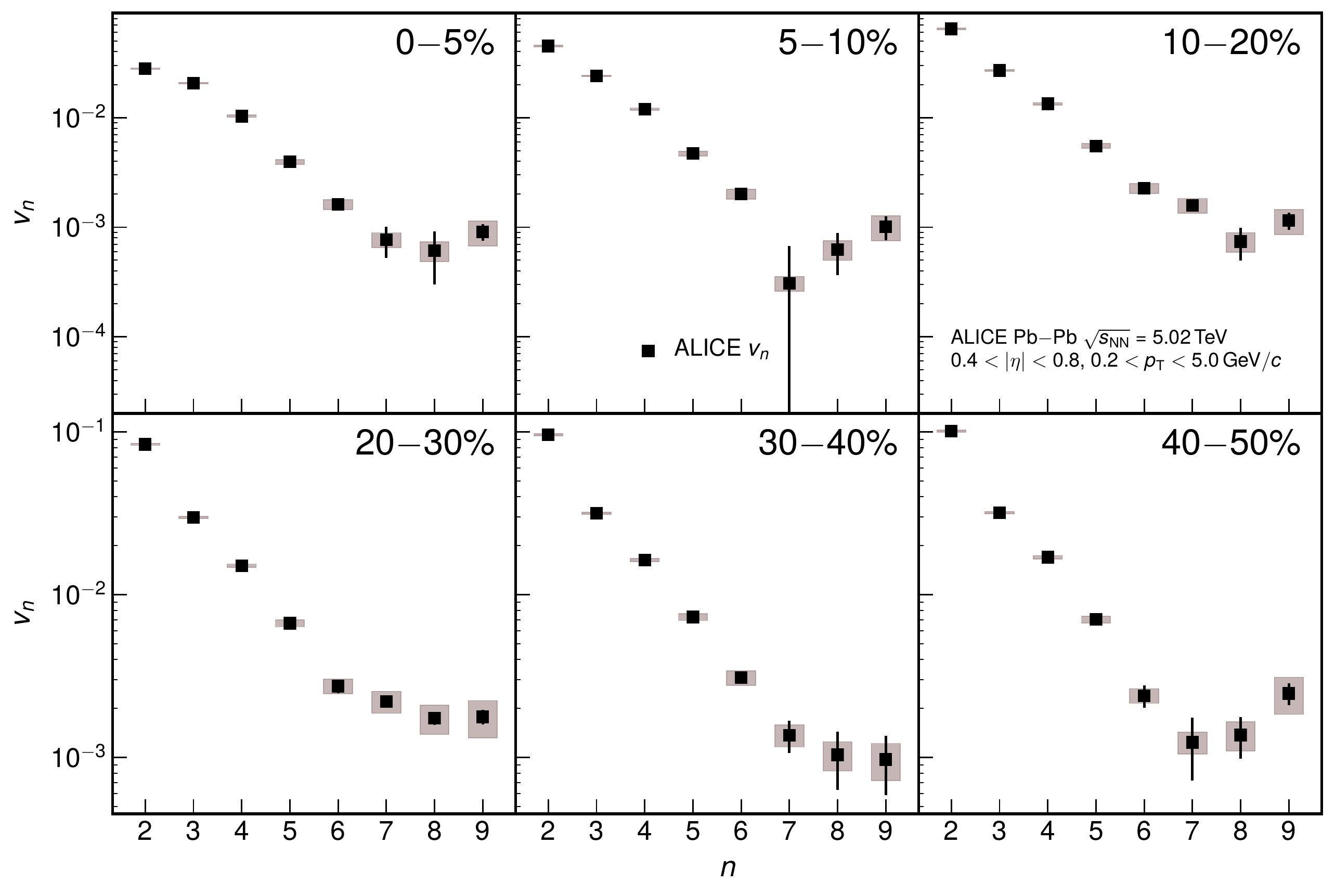}
	\caption{$v_n$ as a function of the harmonic order $n$ for various centrality intervals.}
 	\label{fig:pspectra}
\end{figure}

To study the dependence on the harmonic order of the anisotropy coefficients~\cite{Shuryak:2017aol}, Fig.~\ref{fig:pspectra} shows values of different coefficients as a function of $n$ for all centralities.
This presentation is particularly well suited in visualizing the harmonic dependence, and the acoustic scaling~\cite{Shuryak:2017aol} observed across the harmonic orders.
The decrease in $v_n$ with increasing harmonic order up to $n=7$ indicates viscous damping~\cite{Shuryak:2017aol}. This means that the higher frequency waveform propagating through the medium should get more damped until freeze-out takes place. In~\cite{Staig:2011wj,Lacey:2013is} the viscosity effect is explained as the main contributor to the observed damping. It is speculated, that another driving factor is the phase of the oscillation itself, which also contributes to the magnitude at the time of freeze-out. The measurements show that there is a hint of $v_9>v_8$, as also predicted in the acoustic model~\cite{Shuryak:2017aol}.

\begin{figure}[tbp]
	\centering
	\hspace{-3em}\includegraphics[width=0.80\textwidth]{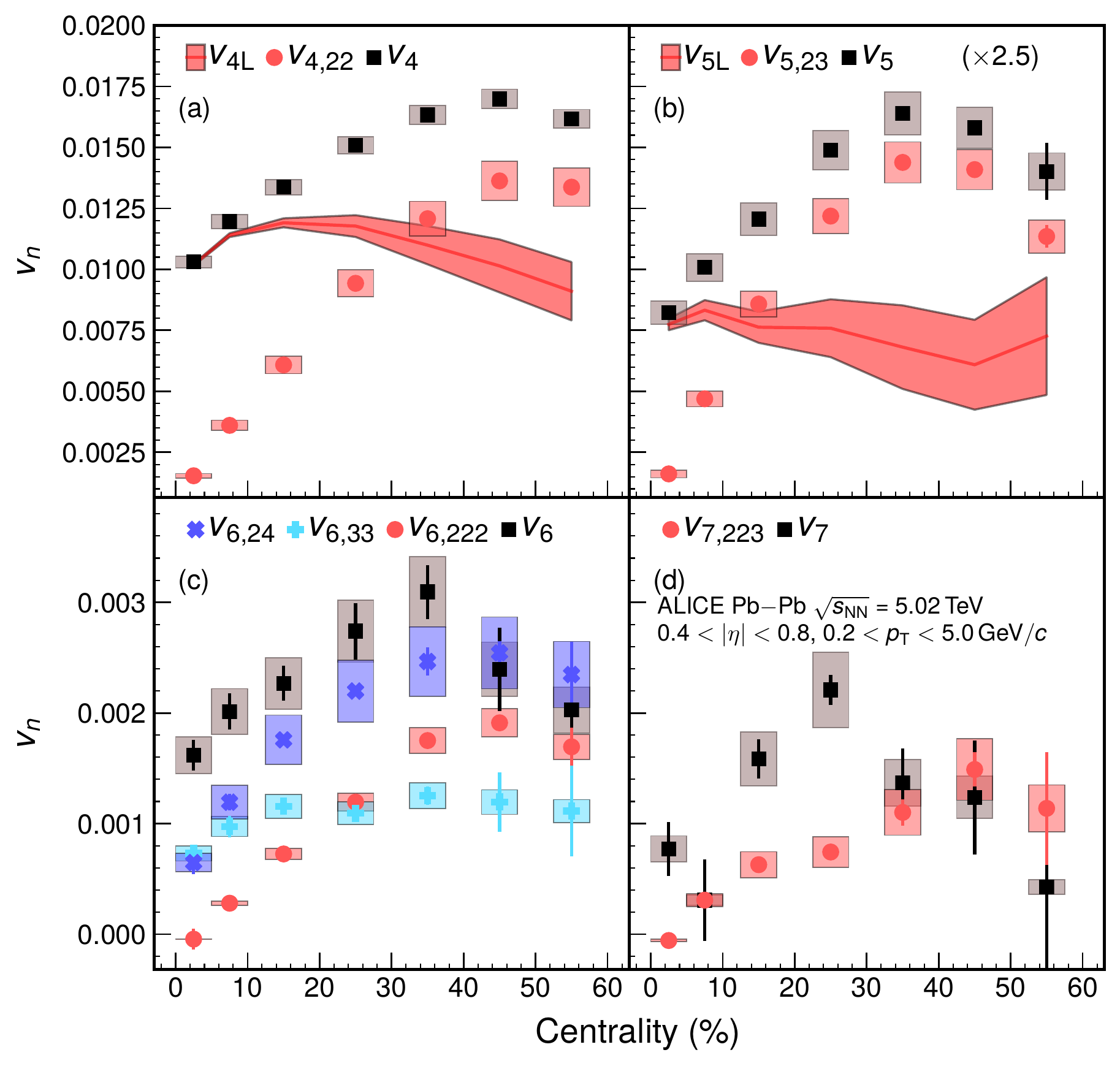}
	\caption{Linear and non-linear flow modes as a function of centrality. The total contribution measured in Pb--Pb collisions at $\sqrt{s_\mathrm{NN}}=5.02\,\mathrm{TeV}$ is shown as black squares. Various non-linear contributions are presented in different red and blue colors, while the linear part, extracted from the aforementioned contributions, is shown as a red band. For panel~(b), the data points are scaled by 2.5 for better visibility across the panels.} 
 	\label{fig:modes}
\end{figure}

Figure~\ref{fig:modes} presents the higher order flow coefficients as well as their linear and non-linear flow modes up to the seventh order as a function of centrality. For the flow harmonics $v_4$ and $v_5$, presented in panels~(a) and (b), respectively, the non-linear contributions are small in central collisions, where the linear contribution is dominant. A weak centrality dependence is observed for the linear component. In case of $v_4$, significant contributions from the second order arise in less central collisions. The $v_5$ coefficient, on the other hand, is largely induced by the low order $v_2$ and $v_3$, indicated by the large $v_{5,23}$. 

Panels (c) and (d) of Fig.~\ref{fig:modes} show the flow modes of $v_6$ and $v_7$. Only the non-linear flow modes of $v_6$ and $v_7$ are presented. The $v_{6,222}$ increases from zero to approximately half of the total $v_6$ in mid-central collisions, while the other mode, $v_{6,33}$, has a much weaker centrality dependence. The relatively large magnitude of these flow modes imply strong contributions from the second and third order harmonics. Finally, $v_{6,24}$ follows the trend of the total magnitude. The magnitude of $v_{6,24}$ comes close to the total, which in turn suggests not only strong contributions from the second harmonic order, but also the fourth one. The $v_{6,24}$ induced by the fourth order is seen to be the dominant contribution to the sixth order from 10\% centrality classes and higher. For the seventh order $v_7$, there are three non-linear contributions, of which $v_{7,223}$ is measured. The centrality dependence is similar as with the $v_6$ coefficient, and there is a similar general trend as for the lower order harmonics among the non-linear flow modes.


\begin{figure}[tbp]
	\centering
	\includegraphics[width=1.0\textwidth]{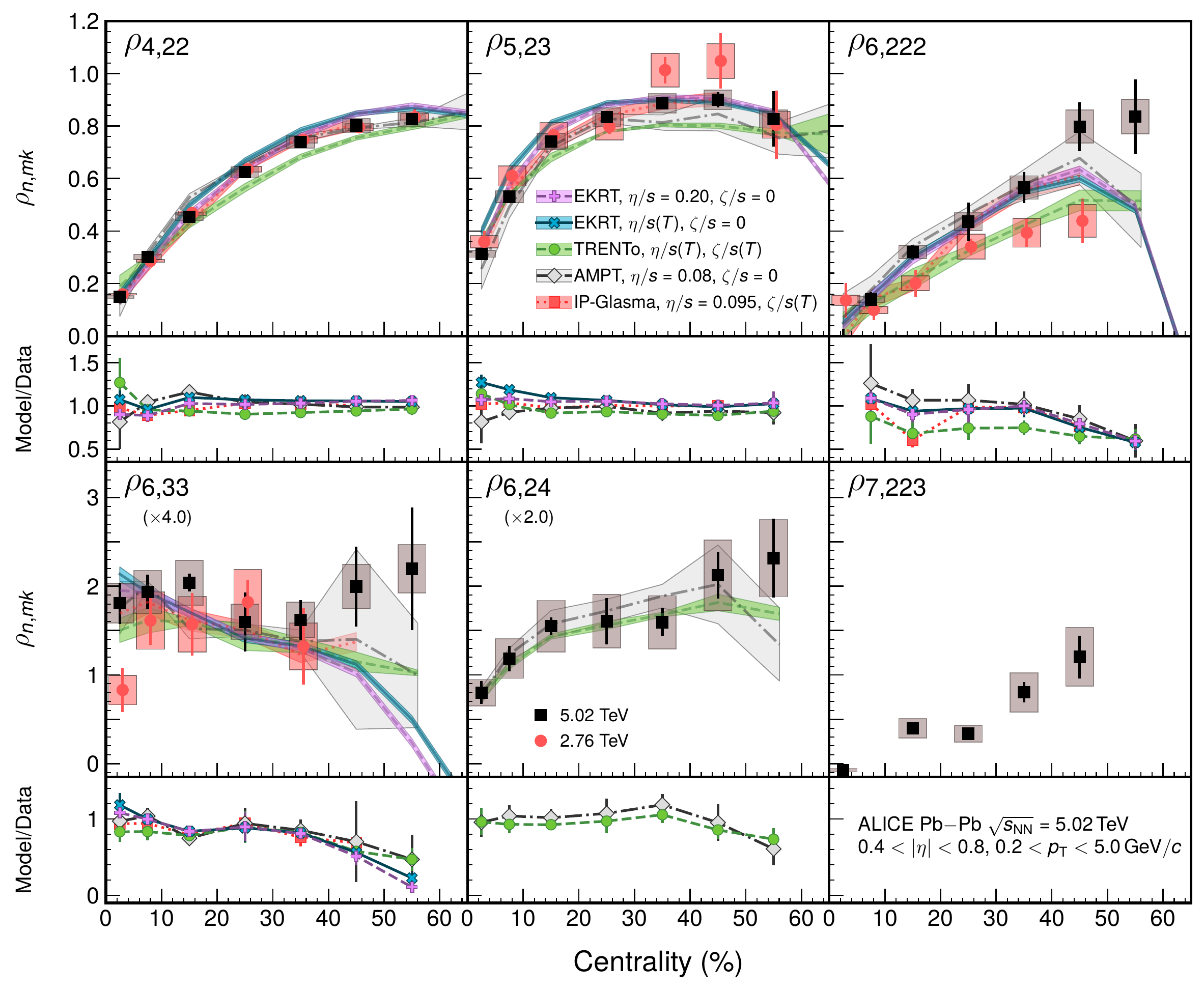}
	\caption{Symmetry-plane correlations as a function of centrality in Pb--Pb collisions at $\sqrt{s_\mathrm{NN}}=5.02\,\mathrm{TeV}$ (black markers) compared with those in Pb--Pb collisions at $\sqrt{s_\mathrm{NN}}=2.76\,\mathrm{TeV}$~\cite{Acharya:2017zfg}, along with five different hydrodynamic calculations shown as color bands. On the bottom part of each panel, the ratios between model calculations and the data are shown. For some panels, the data points are scaled by an indicated factor for better visibility.} 
 	\label{fig:rho}
\end{figure}

The coefficients $\rho_{n,mk}$, quantifying the correlations amongst different symmetry planes, are presented as a function of centrality in Fig.~\ref{fig:rho}. Except for $\rho_{6,33}$, all coefficients indicate an increase in correlation between symmetry planes with increasing centrality class of the collision. The measurements generally agree with the ones obtained at the lower energy. The $\rho_{6,222}$ is the only coefficient for which an energy dependence can be observed. 
The hydrodynamic calculations reproduce the measurements within the large theoretical uncertainties. For $\rho_{4,22}$, $\rho_{5,23}$, and $\rho_{6,222}$, {\trento+\allowbreak{}param3} however underestimates the data in mid-central collisions. 

\begin{figure}[tbp]
	\centering
	\includegraphics[width=1.0\textwidth]{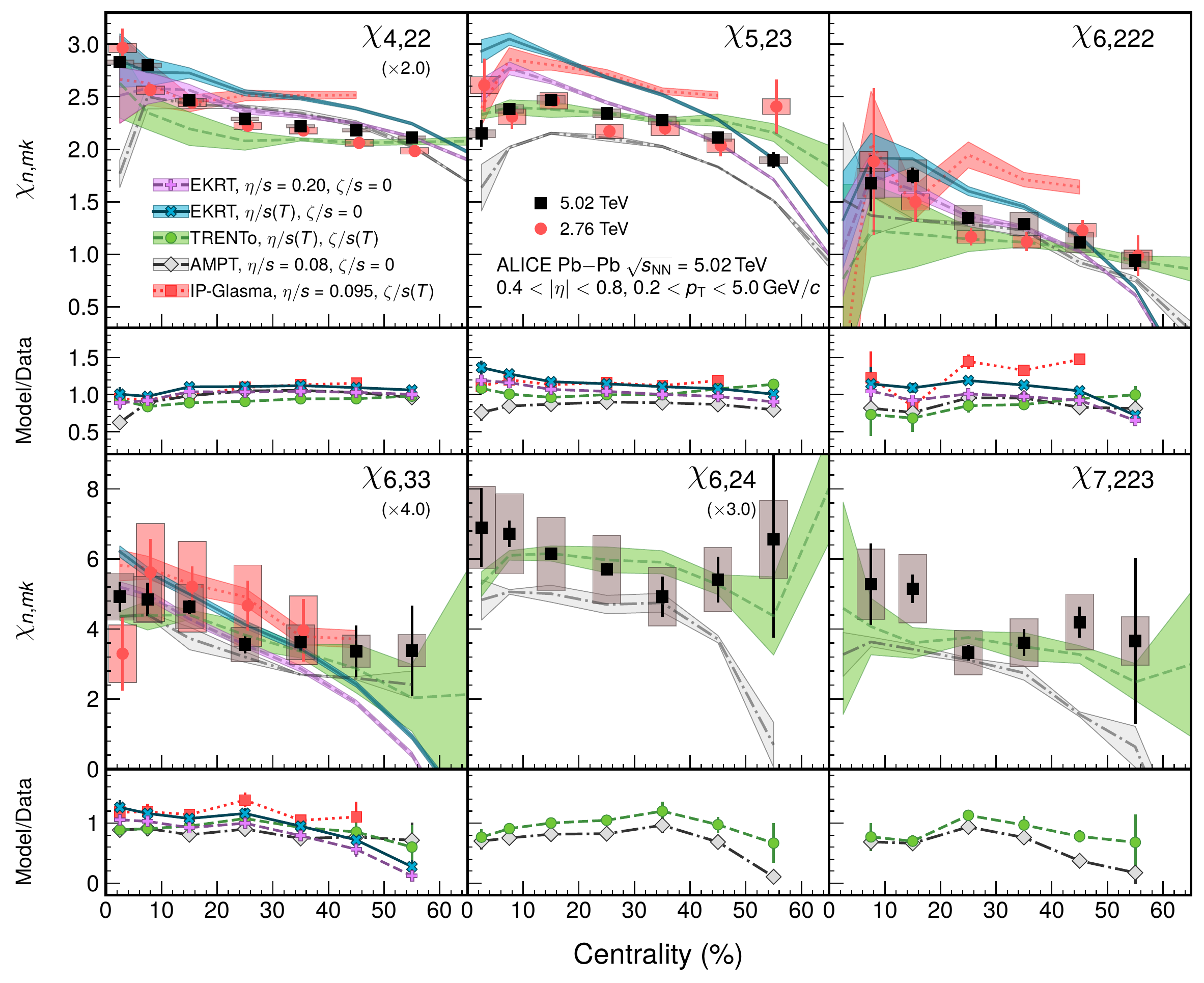}
	\caption{Non-linear mode coefficients as a function of centrality in Pb--Pb collisions at $\sqrt{s_\mathrm{NN}}=5.02\,\mathrm{TeV}$ (black markers) compared with those from $\sqrt{s_\mathrm{NN}}=2.76\,\mathrm{TeV}$ (red markers)~\cite{Acharya:2017zfg}, along with five different hydrodynamic calculations shown as color bands. On the bottom part of each panel, the ratios between model calculations and the data are shown. For some panels, the points are scaled by an indicated factor for better visibility across the panels.}
 	\label{fig:chi}
\end{figure}

Finally, the non-linear flow mode coefficients are presented in Fig.~\ref{fig:chi}. Six coefficients are measured, of which four are compared with the lower beam energy results available in~\cite{Acharya:2017zfg}. For $\chi_{4,22}$ and $\chi_{5,23}$, the centrality dependence and overall magnitude agree well with the results from the lower beam energy. The centrality dependence of the new data is similar to the previous results: a larger value in more central collisions, decreasing close to unity towards 50\% centrality. 

All of the non-linear flow mode coefficients for the sixth harmonic agree with the previous measurements. The centrality dependence of $\chi_{6,222}$ is similar to the ones of the lower order coefficients, and the overall magnitude similar to $\chi_{4,22}$. As for $\chi_{6,33}$, no clear centrality dependence is observed within the current experimental uncertainties. Whereas the previous measurements are unable to distinguish between the magnitudes of $\chi_{6,222}$ and $\chi_{6,33}$, the current results show that $\chi_{6,222}>\chi_{6,33}$ across the whole centrality interval. 
For $\chi_{7,223}$, the overall magnitude is larger than for the other non-linear flow mode coefficients.

The hydrodynamic calculations for the non-linear flow mode coefficients show slightly more variation compared to the symmetry-plane correlations. As seen from the panels of Fig.~\ref{fig:chi}, one observes the reproduction of the data points by {EKRT+\allowbreak{}param0} up to the modes of the sixth harmonic, and {\trento+\allowbreak{}param3} in all harmonics. 
The {EKRT+\allowbreak{}param1} calculations slightly overestimate the centrality dependence of the non-linear flow mode coefficients. It can be seen that the parameterizations of the EKRT presented here imply $\chi_{n,mk}$ across all harmonic orders to have sensitivity to $\eta/s$, whereas in the previous calculations in~\cite{Acharya:2017zfg}, weak $\eta/s$ dependence was found for $\chi_{4,22}$ and $\chi_{6,222}$.
The fifth order coefficient $\chi_{5,23}$ is expected to be quite sensitive to $\eta/s$ in central collisions as can be seen from the difference of the predicted values from {EKRT+\allowbreak{}param0} and {EKRT+\allowbreak{}param1}. 
The {AMPT+\allowbreak{}param2} calculations underestimate the magnitude of some of the measured non-linear flow mode coefficients in various centrality classes, especially $\chi_{5,23}$, $\chi_{7,223}$ as well as $\chi_{6,24}$. The {IP-Glasma+\allowbreak{}param4} calculations overestimate the measurements in some centrality intervals.

\begin{figure}[tbp]
	\centering
	\includegraphics[width=1.0\textwidth]{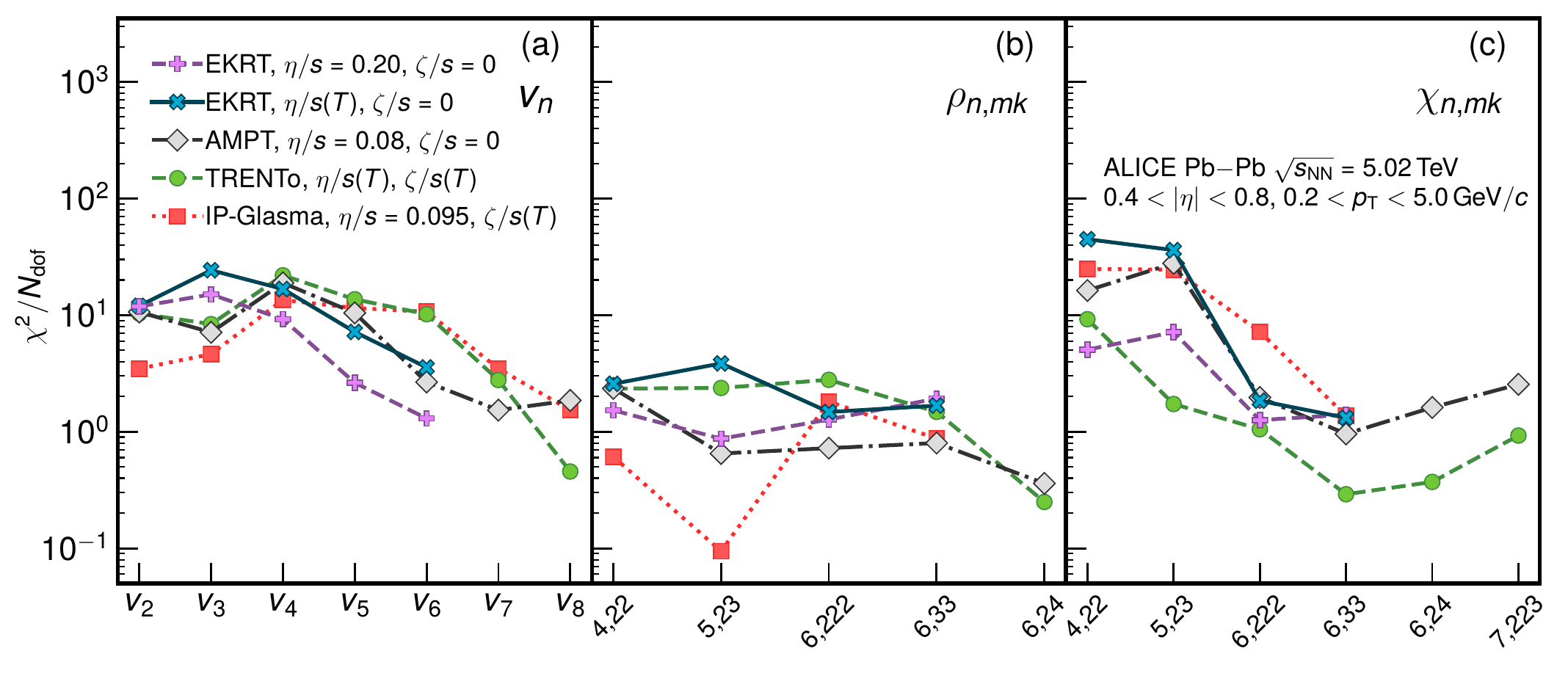}
	\caption{Overview of various model comparisons with data, quantified by $\chi^2/N_\mathrm{dof}$. Lower $\chi^2/N_\mathrm{dof}$ represents a better overall description for a given observable.}
 	\label{fig:chisq}
\end{figure}

The agreement between data and the model calculations is quantified by calculating the reduced $\chi^2/N_\mathrm{dof}$ defined as
\begin{equation}
\chi^2/N_\mathrm{dof}=\frac{1}{N_\mathrm{dof}}\sum_{i=1}^{N_\mathrm{dof}}\frac{(y_i-f_i)^2}{\sigma_i^2},
\end{equation}
where $y_i$ and $f_i$ are the values for data and calculations, respectively, and $\sigma_i^2=\sigma_{i,\mathrm{stat}}^2+\sigma_{i,\mathrm{syst}}^2+\sigma_{f_i,\mathrm{stat}}^2$ is the quadratic uncertainty in terms of statistical measurement $\sigma_{i,\mathrm{stat}}$, model uncertainties $\sigma_{f_i,\mathrm{stat}}$, and systematic uncertainties $\sigma_{i,\mathrm{syst}}$ in centrality bin $i$. Here $N_\mathrm{dof}$ represents the number of data points across the centrality interval.

The $\chi^2/N_\mathrm{dof}$ for the flow coefficients are presented in Fig.~\ref{fig:chisq}, panel~(a). It is observed that {IP-Glasma+\allowbreak{}param4} gives the best description of $v_2$ and $v_3$ compared to the other models, indicated by the overall low value of $\chi^2/N_\mathrm{dof}$. However, the overall performance of {IP-Glasma+\allowbreak{}param4} is considerably worse at $n\geq 4$, for which the data are overestimated, as seen in Fig.~\ref{fig:vns}. For $v_4$ to $v_6$, {EKRT+\allowbreak{}param0} gives the lowest value of $\chi^2/N_\mathrm{dof}$. In the case of {EKRT+\allowbreak{}param1}, the $\chi^2/N_\mathrm{dof}$ is slightly worse than {EKRT+\allowbreak{}param0}. The $\chi^2/N_\mathrm{dof}$ of {\trento+\allowbreak{}param3} is very close to that of {IP-Glasma+\allowbreak{}param4}, indicating a comparable description of data between the two model configurations. At low harmonic orders, {\trento+\allowbreak{}param3} performs slightly worse than {IP-Glasma+\allowbreak{}param4}. For $n\geq 4$, description of the data between these two models are comparable except for $n=8$, where {\trento+\allowbreak{}param3} clearly has a better magnitude and centrality depiction. Notably this can be seen for $v_8$ where the $\chi^2/N_\mathrm{dof}$ value is the lowest across all models. Finally, the performance of {AMPT+\allowbreak{}param2} can be considered good within the reported $\chi^2/N_\mathrm{dof}$ values.
It is noted that the magnitude of $v_7$ is best depicted by {AMPT+\allowbreak{}param2} amongst the three models used.

The performance of the models with respect to the symmetry-plane correlations is quantified in panel~(b) of Fig.~\ref{fig:chisq}. {IP-Glasma+\allowbreak{}param4} has by far the best estimates of $\rho_{n,mk}$ for $\rho_{4,22}$ and $\rho_{5,23}$. For other models, the model depiction is comparable. In low harmonic orders, {EKRT+\allowbreak{}param0} shows good agreement with the data, as well as {AMPT+\allowbreak{}param2}, which has the best agreement in higher harmonic orders. For {\trento+\allowbreak{}param3}, the agreement is slightly worse for $\rho_{5,23}$ and $\rho_{6,222}$. 

The panel~(c) of Fig.~\ref{fig:chisq} shows the $\chi^2/N_\mathrm{dof}$ for non-linear flow mode coefficients. As seen also in Fig.~\ref{fig:chi}, {\trento+\allowbreak{}param3} consistently provides the most successful overall description of the data. For other models the data are more frequently over- or underestimated. 
{\trento+\allowbreak{}param3} estimates $\chi_{n,mk}$ better than it does the $v_n$ coefficients, for which significant overestimation was present at almost every harmonic order (see Fig.~\ref{fig:vns}). For {EKRT+\allowbreak{}param0} the agreement is good, but the calculation over- or underestimates in some cases  especially in most central or mid-peripheral collisions.
Most of the observables are better described by the calculations using {EKRT+\allowbreak{}param0} with a const $\eta/s=0.2$ as compared to results from {EKRT+\allowbreak{}param1} which uses a temperature dependent $\eta/s$ value.
{AMPT+\allowbreak{}param2} performs worse for low-order harmonics as it overpredicts the data in central and mid-central collisions. Of the five configurations, {IP-Glasma+\allowbreak{}param4} describes the data worse in all harmonic orders.

The deviation of the calculated results from the measured value of each observable is of the same order of magnitude for the different models.
Even where the model results show gross agreement with overall features in data, the values of $\chi^2/N_\mathrm{dof}$ vary considerably from one harmonic order to another. Considering the $\chi^2/N_\mathrm{dof}$ to be a goodness-of-fit estimate to validate any model, these variations suggest that the sensitivity of the different observables on the initial conditions, $\eta/s$, and $\zeta/s$ are reflected differently in the model calculations. Since the current uncertainties in the model calculations are large for higher order harmonics, the absolute $\chi^2$ test should not be over-interpreted. Both, improved statistical uncertainties in the model calculations and different values of input parameters would be beneficial. However, large sets of calculations in many parameter spaces require substantial computing power. In order to constrain the model parameters Bayesian analysis can provide a plausible approach as demonstrated in~\cite{Bernhard:2016tnd,Bernhard2019}. At present it is limited to low harmonic-order observables, and the extracted parameters have large uncertainties. Extending the Bayesian analysis to include the results in this paper will help reduce the uncertainties of the model parameters. 
 

\section{Summary}

The measurements of anisotropic flow coefficients ($v_n$), non-linear flow mode coefficients ($\chi_{n,mk}$), and correlations among different symmetry planes ($\rho_{n,mk}$) in Pb--Pb collisions at $\sqrt{s_\mathrm{NN}}=5.02\,\mathrm{TeV}$ are presented. 
The anisotropic flow coefficients are measured up to $v_9$, where $v_8$ and $v_9$ are measured for the first time at LHC energies. It is observed that $v_n$ decreases as $n$ increases, observing $n$-ordered damping up to $n=7$. The $v_n$ is found to be enhanced for $n>7$. The non-linear contribution becomes dominant towards peripheral collisions in all harmonic orders. The strength of the non-linear flow mode and the symmetry-plane correlations depends also on harmonic orders. The non-linear flow mode coefficients show a clear centrality and harmonic order dependencies and the strongest non-linear mode coefficients is observed for the fifth and seventh harmonic orders. 

These results are compared with various hydrodynamic model calculations with different initial conditions, as well as different pa\-{}ra\-{}me\-teri\-za\-tions of $\eta/s$ and $\zeta/s$. None of the models presented in this paper simultaneously describe the $v_n$ coefficients, $\chi_{n,mk}$, or $\rho_{n,mk}$. Based on the model and data comparisons, among all the models, the event-by-event viscous hydrodynamic model with EKRT initial conditions and a constant $\eta/s=0.2$ is observed to describe the data best, as far as the harmonics up to the sixth order are concerned. As a result further tuning is required to find the accurate parameterization of $\eta/s$ and $\zeta/s$. It is found that the different order harmonic observables presented in this paper have different sensitivities to the initial conditions and the system properties. These results allow further model parameters to be optimized and the initial conditions and the transport properties of nuclear matter in ultra-relativistic heavy-ion collisions to be better constrained.

%
%

\newenvironment{acknowledgement}{\relax}{\relax}
\begin{acknowledgement}
\section*{Acknowledgements}

The ALICE Collaboration would like to thank all its engineers and technicians for their invaluable contributions to the construction of the experiment and the CERN accelerator teams for the outstanding performance of the LHC complex.
The ALICE Collaboration gratefully acknowledges the resources and support provided by all Grid centres and the Worldwide LHC Computing Grid (WLCG) collaboration.
The ALICE Collaboration acknowledges the following funding agencies for their support in building and running the ALICE detector:
A. I. Alikhanyan National Science Laboratory (Yerevan Physics Institute) Foundation (ANSL), State Committee of Science and World Federation of Scientists (WFS), Armenia;
Austrian Academy of Sciences, Austrian Science Fund (FWF): [M 2467-N36] and Nationalstiftung f\"{u}r Forschung, Technologie und Entwicklung, Austria;
Ministry of Communications and High Technologies, National Nuclear Research Center, Azerbaijan;
Conselho Nacional de Desenvolvimento Cient\'{\i}fico e Tecnol\'{o}gico (CNPq), Financiadora de Estudos e Projetos (Finep), Funda\c{c}\~{a}o de Amparo \`{a} Pesquisa do Estado de S\~{a}o Paulo (FAPESP) and Universidade Federal do Rio Grande do Sul (UFRGS), Brazil;
Ministry of Education of China (MOEC) , Ministry of Science \& Technology of China (MSTC) and National Natural Science Foundation of China (NSFC), China;
Ministry of Science and Education and Croatian Science Foundation, Croatia;
Centro de Aplicaciones Tecnol\'{o}gicas y Desarrollo Nuclear (CEADEN), Cubaenerg\'{\i}a, Cuba;
Ministry of Education, Youth and Sports of the Czech Republic, Czech Republic;
The Danish Council for Independent Research | Natural Sciences, the VILLUM FONDEN and Danish National Research Foundation (DNRF), Denmark;
Helsinki Institute of Physics (HIP), Finland;
Commissariat \`{a} l'Energie Atomique (CEA), Institut National de Physique Nucl\'{e}aire et de Physique des Particules (IN2P3) and Centre National de la Recherche Scientifique (CNRS) and R\'{e}gion des  Pays de la Loire, France;
Bundesministerium f\"{u}r Bildung und Forschung (BMBF) and GSI Helmholtzzentrum f\"{u}r Schwerionenforschung GmbH, Germany;
General Secretariat for Research and Technology, Ministry of Education, Research and Religions, Greece;
National Research, Development and Innovation Office, Hungary;
Department of Atomic Energy Government of India (DAE), Department of Science and Technology, Government of India (DST), University Grants Commission, Government of India (UGC) and Council of Scientific and Industrial Research (CSIR), India;
Indonesian Institute of Science, Indonesia;
Centro Fermi - Museo Storico della Fisica e Centro Studi e Ricerche Enrico Fermi and Istituto Nazionale di Fisica Nucleare (INFN), Italy;
Institute for Innovative Science and Technology , Nagasaki Institute of Applied Science (IIST), Japanese Ministry of Education, Culture, Sports, Science and Technology (MEXT) and Japan Society for the Promotion of Science (JSPS) KAKENHI, Japan;
Consejo Nacional de Ciencia (CONACYT) y Tecnolog\'{i}a, through Fondo de Cooperaci\'{o}n Internacional en Ciencia y Tecnolog\'{i}a (FONCICYT) and Direcci\'{o}n General de Asuntos del Personal Academico (DGAPA), Mexico;
Nederlandse Organisatie voor Wetenschappelijk Onderzoek (NWO), Netherlands;
The Research Council of Norway, Norway;
Commission on Science and Technology for Sustainable Development in the South (COMSATS), Pakistan;
Pontificia Universidad Cat\'{o}lica del Per\'{u}, Peru;
Ministry of Science and Higher Education and National Science Centre, Poland;
Korea Institute of Science and Technology Information and National Research Foundation of Korea (NRF), Republic of Korea;
Ministry of Education and Scientific Research, Institute of Atomic Physics and Ministry of Research and Innovation and Institute of Atomic Physics, Romania;
Joint Institute for Nuclear Research (JINR), Ministry of Education and Science of the Russian Federation, National Research Centre Kurchatov Institute, Russian Science Foundation and Russian Foundation for Basic Research, Russia;
Ministry of Education, Science, Research and Sport of the Slovak Republic, Slovakia;
National Research Foundation of South Africa, South Africa;
Swedish Research Council (VR) and Knut \& Alice Wallenberg Foundation (KAW), Sweden;
European Organization for Nuclear Research, Switzerland;
Suranaree University of Technology (SUT), National Science and Technology Development Agency (NSDTA) and Office of the Higher Education Commission under NRU project of Thailand, Thailand;
Turkish Atomic Energy Agency (TAEK), Turkey;
National Academy of  Sciences of Ukraine, Ukraine;
Science and Technology Facilities Council (STFC), United Kingdom;
National Science Foundation of the United States of America (NSF) and United States Department of Energy, Office of Nuclear Physics (DOE NP), United States of America.    
\end{acknowledgement}

\bibliographystyle{utphys}   
\bibliography{references}
\newpage
\appendix
\section{List of Observables}
\label{sec:obslist}
In this section the complete list of the measured observables is presented. By root-mean-squaring the equations in Eq.~(\ref{eq:nldecomp}), one obtains a starting point for the definitions presented in this section. Provided that the linear and non-linear parts are uncorrelated, the following harmonic projections are obtained
\begin{equation}
\label{eq:proj2}
\begin{alignedat}{3}
	v_{4,22}&=\frac{\Re\langle V_4 (V_2^*)^2\rangle}{\sqrt{\langle|V_2|^4\rangle}}\quad
	&v_{5,23}=\frac{\Re\langle V_5 V_2^* V_3^*\rangle}{\sqrt{\langle|V_2|^2|V_3|^2\rangle}}\\
	&\approx \langle v_4\cos(4\psi_4-4\psi_2)\rangle,\quad&\approx\langle v_5\cos(5\psi_5-3\psi_3-2\psi_2)\rangle,\\
	v_{6,222}&=\frac{\Re\langle V_6 (V_2^*)^3\rangle}{\sqrt{\langle|V_2|^6\rangle}}\quad
	&v_{6,24}=\frac{\Re\langle V_6 V_2^* V_4^*\rangle}{\sqrt{\langle|V_2|^2|V_4|^2\rangle}}\\
	&\approx \langle v_6\cos(6\psi_6-6\psi_2)\rangle,\quad&\approx\langle v_6\cos(6\psi_6-4\psi_4-2\psi_2)\rangle,\\
	v_{6,33}&=\frac{\Re\langle V_6 (V_3^*)^2\rangle}{\sqrt{\langle|V_3|^4\rangle}}\quad
	&v_{7,223}=\frac{\Re\langle V_7 (V_2^*)^2 V_3^*\rangle}{\sqrt{\langle|V_2|^4|V_3|^2\rangle}}\\
	&\approx\langle v_6\cos(6\psi_6-6\psi_3)\rangle,\quad&\approx\langle v_7\cos(7\psi_7-4\psi_2-3\psi_3)\rangle,\\
	%
	%
	v_{8,233}&=\frac{\Re\langle V_8 V_2^* (V_3^*)^2\rangle}{\sqrt{\langle|V_2|^2|V_3|^4\rangle}}\\
	&\approx\langle v_8\cos(8\psi_8-2\psi_2-6\psi_3)\rangle,\quad&
\end{alignedat}
\end{equation}
with $v_{4,22}^2=\chi_{4,22}^2\langle|V_2|^4\rangle$, $v_{5,23}^2=\chi_{5,23}^2\langle|V_2|^2|V_3|^2\rangle$, \dots.
The rest of the observables we define using the harmonic projections in Eq.~(\ref{eq:proj2}). The symmetry plane correlations are defined as
\begin{equation}
\begin{alignedat}{3}
	\rho_{4,22}&=\frac{v_{4,22}}{v_4},\quad
	\rho_{5,23}&=\frac{v_{5,23}}{v_5},\\
	\rho_{6,222}&=\frac{v_{6,222}}{v_6},\quad
	\rho_{7,223}&=\frac{v_{7,334}}{v_7},\\
	\rho_{6,33}&=\frac{v_{6,33}}{v_6},
\end{alignedat}
\end{equation}
and the non-linear mode coefficients
\begin{equation}
\begin{split}
&\quad\quad\begin{alignedat}{3}
	\chi_{4,22}&=\frac{v_{4,22}}{\sqrt{\langle v_2^4\rangle}},\quad
	\chi_{5,23}&=\frac{v_{5,23}}{\sqrt{\langle v_2^2 v_3^2\rangle}},\\
	\chi_{6,222}&=\frac{v_{6,222}}{\sqrt{\langle v_2^6\rangle}},\quad
	\chi_{7,223}&=\frac{v_{7,223}}{\sqrt{\langle v_2^4 v_3^2\rangle}},\\
	\chi_{6,33}&=\frac{v_{6,33}}{\sqrt{\langle v_3^4\rangle}},
\end{alignedat}\\
&\begin{split}
	\chi_{6,24}=\Re&\frac{\langle V_6 V_2^* V_4^*\rangle\langle v_2^4\rangle-\langle V_6 (V_2^*)^3\rangle\langle V_4 (V_2^*)^2\rangle}{(\langle v_4^2\rangle\langle v_2^4\rangle-\langle V_4 (V_2^*)^2\rangle^2)\langle v_2^2\rangle}.
\end{split}
\end{split}
\end{equation}

The higher order superpositions in Eq.~(\ref{eq:nldecomp}) include the coupling constants for the higher order linear responses. In a more complete treatment \cite{Qian:2016fpi}, the extraction of the higher order non-linear flow mode coefficients are performed by correlating the corresponding superpositions with those of the relevant harmonics, effectively resulting in a more general projection. The results agree with the expressions in Eq.~(\ref{eq:chi}), and generate additional high order linear coupling coefficients
\begin{equation}
\label{eq:chilin}
	\begin{split}
	\chi_{6,24}=\Re&\frac{\langle V_6 V_2^* V_4^*\rangle\langle v_2^4\rangle-\langle V_6 (V_2^*)^3\rangle\langle V_4 (V_2^*)^2\rangle}{(\langle v_4^2\rangle\langle v_2^4\rangle-\langle V_4 (V_2^*)^2\rangle^2)\langle v_2^2\rangle},\\
	\chi_{7,25}=\Re&\frac{\langle V_7 V_2^* V_5^*\rangle\langle v_2^2 v_3^2\rangle-\langle V_7 (V_2^*)^2 V_3^*\rangle\langle V_5 V_2^* V_3^*\rangle}{(\langle v_5^2\rangle\langle v_2^2 v_3^2\rangle-\langle V_5 V_2^* V_3^*\rangle^2)\langle v_2^2\rangle},\\
	\chi_{7,34}=\Re&\frac{\langle V_7 V_3^* V_4^*\rangle\langle v_2^4\rangle-\langle V_7 (V_2^*)^2 V_3^*\rangle\langle V_4 (V_2^*)^2\rangle}{(\langle v_4^2\rangle\langle v_2^4\rangle-\langle V_4 (V_2^*)^2\rangle^2)\langle v_3^2\rangle}.
	\end{split}
\end{equation}

\newpage
\section{The ALICE Collaboration}
\label{app:collab}

\begingroup
\small
\begin{flushleft}
S.~Acharya\Irefn{org141}\And 
D.~Adamov\'{a}\Irefn{org94}\And 
A.~Adler\Irefn{org74}\And 
J.~Adolfsson\Irefn{org80}\And 
M.M.~Aggarwal\Irefn{org99}\And 
G.~Aglieri Rinella\Irefn{org33}\And 
M.~Agnello\Irefn{org30}\And 
N.~Agrawal\Irefn{org10}\textsuperscript{,}\Irefn{org53}\And 
Z.~Ahammed\Irefn{org141}\And 
S.~Ahmad\Irefn{org16}
% ...
\And
\renewcommand\labelenumi{\textsuperscript{\theenumi}~}

\section*{Affiliation notes}
\renewcommand\theenumi{\roman{enumi}}
\begin{Authlist}
\item \Adef{org*}Deceased
\item \Adef{orgI}Italian National Agency for New Technologies, Energy and Sustainable Economic Development (ENEA), Bologna, Italy
\item \Adef{orgII}Dipartimento DET del Politecnico di Torino, Turin, Italy
\item \Adef{orgIII}M.V. Lomonosov Moscow State University, D.V. Skobeltsyn Institute of Nuclear, Physics, Moscow, Russia
\item \Adef{orgIV}Department of Applied Physics, Aligarh Muslim University, Aligarh, India
\item \Adef{orgV}Institute of Theoretical Physics, University of Wroclaw, Poland
\end{Authlist}

\section*{Collaboration Institutes}
\renewcommand\theenumi{\arabic{enumi}~}
\begin{Authlist}
\item \Idef{org1}A.I. Alikhanyan National Science Laboratory (Yerevan Physics Institute) Foundation, Yerevan, Armenia
\item \Idef{org2}Bogolyubov Institute for Theoretical Physics, National Academy of Sciences of Ukraine, Kiev, Ukraine
\item \Idef{org3}Bose Institute, Department of Physics  and Centre for Astroparticle Physics and Space Science (CAPSS), Kolkata, India
\item \Idef{org4}Budker Institute for Nuclear Physics, Novosibirsk, Russia
\item \Idef{org5}California Polytechnic State University, San Luis Obispo, California, United States
\item \Idef{org6}Central China Normal University, Wuhan, China
\item \Idef{org7}Centre de Calcul de l'IN2P3, Villeurbanne, Lyon, France
\item \Idef{org8}Centro de Aplicaciones Tecnol\'{o}gicas y Desarrollo Nuclear (CEADEN), Havana, Cuba
\item \Idef{org9}Centro de Investigaci\'{o}n y de Estudios Avanzados (CINVESTAV), Mexico City and M\'{e}rida, Mexico
\item \Idef{org10}Centro Fermi - Museo Storico della Fisica e Centro Studi e Ricerche ``Enrico Fermi', Rome, Italy
\item \Idef{org11}Chicago State University, Chicago, Illinois, United States
\item \Idef{org12}China Institute of Atomic Energy, Beijing, China
\item \Idef{org13}Comenius University Bratislava, Faculty of Mathematics, Physics and Informatics, Bratislava, Slovakia
\item \Idef{org14}COMSATS University Islamabad, Islamabad, Pakistan
\item \Idef{org15}Creighton University, Omaha, Nebraska, United States
\item \Idef{org16}Department of Physics, Aligarh Muslim University, Aligarh, India
\item \Idef{org17}Department of Physics, Pusan National University, Pusan, Republic of Korea
\item \Idef{org18}Department of Physics, Sejong University, Seoul, Republic of Korea
\item \Idef{org19}Department of Physics, University of California, Berkeley, California, United States
\item \Idef{org20}Department of Physics, University of Oslo, Oslo, Norway
\item \Idef{org21}Department of Physics and Technology, University of Bergen, Bergen, Norway
\item \Idef{org22}Dipartimento di Fisica dell'Universit\`{a} 'La Sapienza' and Sezione INFN, Rome, Italy
\item \Idef{org23}Dipartimento di Fisica dell'Universit\`{a} and Sezione INFN, Cagliari, Italy
\item \Idef{org24}Dipartimento di Fisica dell'Universit\`{a} and Sezione INFN, Trieste, Italy
\item \Idef{org25}Dipartimento di Fisica dell'Universit\`{a} and Sezione INFN, Turin, Italy
\item \Idef{org26}Dipartimento di Fisica e Astronomia dell'Universit\`{a} and Sezione INFN, Bologna, Italy
\item \Idef{org27}Dipartimento di Fisica e Astronomia dell'Universit\`{a} and Sezione INFN, Catania, Italy
\item \Idef{org28}Dipartimento di Fisica e Astronomia dell'Universit\`{a} and Sezione INFN, Padova, Italy
\item \Idef{org29}Dipartimento di Fisica `E.R.~Caianiello' dell'Universit\`{a} and Gruppo Collegato INFN, Salerno, Italy
\item \Idef{org30}Dipartimento DISAT del Politecnico and Sezione INFN, Turin, Italy
\item \Idef{org31}Dipartimento di Scienze e Innovazione Tecnologica dell'Universit\`{a} del Piemonte Orientale and INFN Sezione di Torino, Alessandria, Italy
\item \Idef{org32}Dipartimento Interateneo di Fisica `M.~Merlin' and Sezione INFN, Bari, Italy
\item \Idef{org33}European Organization for Nuclear Research (CERN), Geneva, Switzerland
\item \Idef{org34}Faculty of Electrical Engineering, Mechanical Engineering and Naval Architecture, University of Split, Split, Croatia
\item \Idef{org35}Faculty of Engineering and Science, Western Norway University of Applied Sciences, Bergen, Norway
\item \Idef{org36}Faculty of Nuclear Sciences and Physical Engineering, Czech Technical University in Prague, Prague, Czech Republic
\item \Idef{org37}Faculty of Science, P.J.~\v{S}af\'{a}rik University, Ko\v{s}ice, Slovakia
\item \Idef{org38}Frankfurt Institute for Advanced Studies, Johann Wolfgang Goethe-Universit\"{a}t Frankfurt, Frankfurt, Germany
\item \Idef{org39}Fudan University, Shanghai, China
\item \Idef{org40}Gangneung-Wonju National University, Gangneung, Republic of Korea
\item \Idef{org41}Gauhati University, Department of Physics, Guwahati, India
\item \Idef{org42}Helmholtz-Institut f\"{u}r Strahlen- und Kernphysik, Rheinische Friedrich-Wilhelms-Universit\"{a}t Bonn, Bonn, Germany
\item \Idef{org43}Helsinki Institute of Physics (HIP), Helsinki, Finland
\item \Idef{org44}High Energy Physics Group,  Universidad Aut\'{o}noma de Puebla, Puebla, Mexico
\item \Idef{org45}Hiroshima University, Hiroshima, Japan
\item \Idef{org46}Hochschule Worms, Zentrum  f\"{u}r Technologietransfer und Telekommunikation (ZTT), Worms, Germany
\item \Idef{org47}Horia Hulubei National Institute of Physics and Nuclear Engineering, Bucharest, Romania
\item \Idef{org48}Indian Institute of Technology Bombay (IIT), Mumbai, India
\item \Idef{org49}Indian Institute of Technology Indore, Indore, India
\item \Idef{org50}Indonesian Institute of Sciences, Jakarta, Indonesia
\item \Idef{org51}INFN, Laboratori Nazionali di Frascati, Frascati, Italy
\item \Idef{org52}INFN, Sezione di Bari, Bari, Italy
\item \Idef{org53}INFN, Sezione di Bologna, Bologna, Italy
\item \Idef{org54}INFN, Sezione di Cagliari, Cagliari, Italy
\item \Idef{org55}INFN, Sezione di Catania, Catania, Italy
\item \Idef{org56}INFN, Sezione di Padova, Padova, Italy
\item \Idef{org57}INFN, Sezione di Roma, Rome, Italy
\item \Idef{org58}INFN, Sezione di Torino, Turin, Italy
\item \Idef{org59}INFN, Sezione di Trieste, Trieste, Italy
\item \Idef{org60}Inha University, Incheon, Republic of Korea
\item \Idef{org61}Institut de Physique Nucl\'{e}aire d'Orsay (IPNO), Institut National de Physique Nucl\'{e}aire et de Physique des Particules (IN2P3/CNRS), Universit\'{e} de Paris-Sud, Universit\'{e} Paris-Saclay, Orsay, France
\item \Idef{org62}Institute for Nuclear Research, Academy of Sciences, Moscow, Russia
\item \Idef{org63}Institute for Subatomic Physics, Utrecht University/Nikhef, Utrecht, Netherlands
\item \Idef{org64}Institute of Experimental Physics, Slovak Academy of Sciences, Ko\v{s}ice, Slovakia
\item \Idef{org65}Institute of Physics, Homi Bhabha National Institute, Bhubaneswar, India
\item \Idef{org66}Institute of Physics of the Czech Academy of Sciences, Prague, Czech Republic
\item \Idef{org67}Institute of Space Science (ISS), Bucharest, Romania
\item \Idef{org68}Institut f\"{u}r Kernphysik, Johann Wolfgang Goethe-Universit\"{a}t Frankfurt, Frankfurt, Germany
\item \Idef{org69}Instituto de Ciencias Nucleares, Universidad Nacional Aut\'{o}noma de M\'{e}xico, Mexico City, Mexico
\item \Idef{org70}Instituto de F\'{i}sica, Universidade Federal do Rio Grande do Sul (UFRGS), Porto Alegre, Brazil
\item \Idef{org71}Instituto de F\'{\i}sica, Universidad Nacional Aut\'{o}noma de M\'{e}xico, Mexico City, Mexico
\item \Idef{org72}iThemba LABS, National Research Foundation, Somerset West, South Africa
\item \Idef{org73}Jeonbuk National University, Jeonju, Republic of Korea
\item \Idef{org74}Johann-Wolfgang-Goethe Universit\"{a}t Frankfurt Institut f\"{u}r Informatik, Fachbereich Informatik und Mathematik, Frankfurt, Germany
\item \Idef{org75}Joint Institute for Nuclear Research (JINR), Dubna, Russia
\item \Idef{org76}Korea Institute of Science and Technology Information, Daejeon, Republic of Korea
\item \Idef{org77}KTO Karatay University, Konya, Turkey
\item \Idef{org78}Laboratoire de Physique Subatomique et de Cosmologie, Universit\'{e} Grenoble-Alpes, CNRS-IN2P3, Grenoble, France
\item \Idef{org79}Lawrence Berkeley National Laboratory, Berkeley, California, United States
\item \Idef{org80}Lund University Department of Physics, Division of Particle Physics, Lund, Sweden
\item \Idef{org81}Nagasaki Institute of Applied Science, Nagasaki, Japan
\item \Idef{org82}Nara Women{'}s University (NWU), Nara, Japan
\item \Idef{org83}National and Kapodistrian University of Athens, School of Science, Department of Physics , Athens, Greece
\item \Idef{org84}National Centre for Nuclear Research, Warsaw, Poland
\item \Idef{org85}National Institute of Science Education and Research, Homi Bhabha National Institute, Jatni, India
\item \Idef{org86}National Nuclear Research Center, Baku, Azerbaijan
\item \Idef{org87}National Research Centre Kurchatov Institute, Moscow, Russia
\item \Idef{org88}Niels Bohr Institute, University of Copenhagen, Copenhagen, Denmark
\item \Idef{org89}Nikhef, National institute for subatomic physics, Amsterdam, Netherlands
\item \Idef{org90}NRC Kurchatov Institute IHEP, Protvino, Russia
\item \Idef{org91}NRC «Kurchatov Institute»  - ITEP, Moscow, Russia
\item \Idef{org92}NRNU Moscow Engineering Physics Institute, Moscow, Russia
\item \Idef{org93}Nuclear Physics Group, STFC Daresbury Laboratory, Daresbury, United Kingdom
\item \Idef{org94}Nuclear Physics Institute of the Czech Academy of Sciences, \v{R}e\v{z} u Prahy, Czech Republic
\item \Idef{org95}Oak Ridge National Laboratory, Oak Ridge, Tennessee, United States
\item \Idef{org96}Ohio State University, Columbus, Ohio, United States
\item \Idef{org97}Petersburg Nuclear Physics Institute, Gatchina, Russia
\item \Idef{org98}Physics department, Faculty of science, University of Zagreb, Zagreb, Croatia
\item \Idef{org99}Physics Department, Panjab University, Chandigarh, India
\item \Idef{org100}Physics Department, University of Jammu, Jammu, India
\item \Idef{org101}Physics Department, University of Rajasthan, Jaipur, India
\item \Idef{org102}Physikalisches Institut, Eberhard-Karls-Universit\"{a}t T\"{u}bingen, T\"{u}bingen, Germany
\item \Idef{org103}Physikalisches Institut, Ruprecht-Karls-Universit\"{a}t Heidelberg, Heidelberg, Germany
\item \Idef{org104}Physik Department, Technische Universit\"{a}t M\"{u}nchen, Munich, Germany
\item \Idef{org105}Politecnico di Bari, Bari, Italy
\item \Idef{org106}Research Division and ExtreMe Matter Institute EMMI, GSI Helmholtzzentrum f\"ur Schwerionenforschung GmbH, Darmstadt, Germany
\item \Idef{org107}Rudjer Bo\v{s}kovi\'{c} Institute, Zagreb, Croatia
\item \Idef{org108}Russian Federal Nuclear Center (VNIIEF), Sarov, Russia
\item \Idef{org109}Saha Institute of Nuclear Physics, Homi Bhabha National Institute, Kolkata, India
\item \Idef{org110}School of Physics and Astronomy, University of Birmingham, Birmingham, United Kingdom
\item \Idef{org111}Secci\'{o}n F\'{\i}sica, Departamento de Ciencias, Pontificia Universidad Cat\'{o}lica del Per\'{u}, Lima, Peru
\item \Idef{org112}St. Petersburg State University, St. Petersburg, Russia
\item \Idef{org113}Stefan Meyer Institut f\"{u}r Subatomare Physik (SMI), Vienna, Austria
\item \Idef{org114}SUBATECH, IMT Atlantique, Universit\'{e} de Nantes, CNRS-IN2P3, Nantes, France
\item \Idef{org115}Suranaree University of Technology, Nakhon Ratchasima, Thailand
\item \Idef{org116}Technical University of Ko\v{s}ice, Ko\v{s}ice, Slovakia
\item \Idef{org117}Technische Universit\"{a}t M\"{u}nchen, Excellence Cluster 'Universe', Munich, Germany
\item \Idef{org118}The Henryk Niewodniczanski Institute of Nuclear Physics, Polish Academy of Sciences, Cracow, Poland
\item \Idef{org119}The University of Texas at Austin, Austin, Texas, United States
\item \Idef{org120}Universidad Aut\'{o}noma de Sinaloa, Culiac\'{a}n, Mexico
\item \Idef{org121}Universidade de S\~{a}o Paulo (USP), S\~{a}o Paulo, Brazil
\item \Idef{org122}Universidade Estadual de Campinas (UNICAMP), Campinas, Brazil
\item \Idef{org123}Universidade Federal do ABC, Santo Andre, Brazil
\item \Idef{org124}University of Cape Town, Cape Town, South Africa
\item \Idef{org125}University of Houston, Houston, Texas, United States
\item \Idef{org126}University of Jyv\"{a}skyl\"{a}, Jyv\"{a}skyl\"{a}, Finland
\item \Idef{org127}University of Liverpool, Liverpool, United Kingdom
\item \Idef{org128}University of Science and Technology of China, Hefei, China
\item \Idef{org129}University of South-Eastern Norway, Tonsberg, Norway
\item \Idef{org130}University of Tennessee, Knoxville, Tennessee, United States
\item \Idef{org131}University of the Witwatersrand, Johannesburg, South Africa
\item \Idef{org132}University of Tokyo, Tokyo, Japan
\item \Idef{org133}University of Tsukuba, Tsukuba, Japan
\item \Idef{org134}Universit\'{e} Clermont Auvergne, CNRS/IN2P3, LPC, Clermont-Ferrand, France
\item \Idef{org135}Universit\'{e} de Lyon, Universit\'{e} Lyon 1, CNRS/IN2P3, IPN-Lyon, Villeurbanne, Lyon, France
\item \Idef{org136}Universit\'{e} de Strasbourg, CNRS, IPHC UMR 7178, F-67000 Strasbourg, France, Strasbourg, France
\item \Idef{org137}Universit\'{e} Paris-Saclay Centre d'Etudes de Saclay (CEA), IRFU, D\'{e}partment de Physique Nucl\'{e}aire (DPhN), Saclay, France
\item \Idef{org138}Universit\`{a} degli Studi di Foggia, Foggia, Italy
\item \Idef{org139}Universit\`{a} degli Studi di Pavia, Pavia, Italy
\item \Idef{org140}Universit\`{a} di Brescia, Brescia, Italy
\item \Idef{org141}Variable Energy Cyclotron Centre, Homi Bhabha National Institute, Kolkata, India
\item \Idef{org142}Warsaw University of Technology, Warsaw, Poland
\item \Idef{org143}Wayne State University, Detroit, Michigan, United States
\item \Idef{org144}Westf\"{a}lische Wilhelms-Universit\"{a}t M\"{u}nster, Institut f\"{u}r Kernphysik, M\"{u}nster, Germany
\item \Idef{org145}Wigner Research Centre for Physics, Budapest, Hungary
\item \Idef{org146}Yale University, New Haven, Connecticut, United States
\item \Idef{org147}Yonsei University, Seoul, Republic of Korea
\end{Authlist}
\endgroup
\end{document}